\documentclass[10pt,twocolumn,letterpaper]{article}

\usepackage{cvpr}              

\usepackage{graphicx}
\usepackage{amsmath}
\usepackage{amssymb}
\usepackage{booktabs}
\usepackage{enumitem}
\usepackage{multirow}
\usepackage[accsupp]{axessibility}  

\usepackage{algorithm}
\usepackage{algorithmicx}
\usepackage{algpseudocode}
\usepackage{amsmath}

\usepackage[pagebackref,breaklinks,colorlinks]{hyperref}

\usepackage[capitalize]{cleveref}
\Crefname{section}{Section}{Sections}
\crefname{section}{Sec.}{Secs.}
\Crefname{table}{Table}{Tables}
\crefname{table}{Tab.}{Tabs.}

\begin{document}

\title{Learning Trajectory-Aware Transformer for Video Super-Resolution}

\author{Chengxu Liu\textsuperscript{1}\thanks{This work was done while Chengxu Liu was a research intern at Microsoft Research Asia.}, Huan Yang\textsuperscript{2}, Jianlong Fu\textsuperscript{2}, Xueming Qian\textsuperscript{1}\\
\textsuperscript{1}Xi’an Jiaotong University \quad\textsuperscript{2}Microsoft Research Asia\\
{\tt\small liuchx97@gmail.com, \{huayan, jianf\}@microsoft.com, qianxm@mail.xjtu.edu.cn}
}

\maketitle

\begin{abstract}
Video super-resolution (VSR) aims to restore a sequence of high-resolution (HR) frames from their low-resolution (LR) counterparts. Although some progress has been made, there are grand challenges to effectively utilize temporal dependency in entire video sequences. Existing approaches usually align and aggregate video frames from limited adjacent frames (e.g., 5 or 7 frames), which prevents these approaches from satisfactory results. In this paper, we take one step further to enable effective spatio-temporal learning in videos. We propose a novel \textbf{T}rajectory-aware \textbf{T}ransformer for \textbf{V}ideo \textbf{S}uper-\textbf{R}esolution (TTVSR). In particular, we formulate video frames into several pre-aligned trajectories which consist of continuous visual tokens. For a query token, self-attention is only learned on relevant visual tokens along spatio-temporal trajectories. Compared with vanilla vision Transformers, such a design significantly reduces the computational cost and enables Transformers to model long-range features. We further propose a cross-scale feature tokenization module to overcome scale-changing problems that often occur in long-range videos. Experimental results demonstrate the superiority of the proposed TTVSR over state-of-the-art models, by extensive quantitative and qualitative evaluations in four widely-used video super-resolution benchmarks. Both code and pre-trained models can be downloaded at \href{https://github.com/researchmm/TTVSR.git}{https://github.com/researchmm/TTVSR}.
\end{abstract}

\section{Introduction}
\label{sec:intro}
\begin{figure}[h]
  \setlength{\belowcaptionskip}{-0.4cm}
  \centering
  \includegraphics[width=1.0\linewidth]{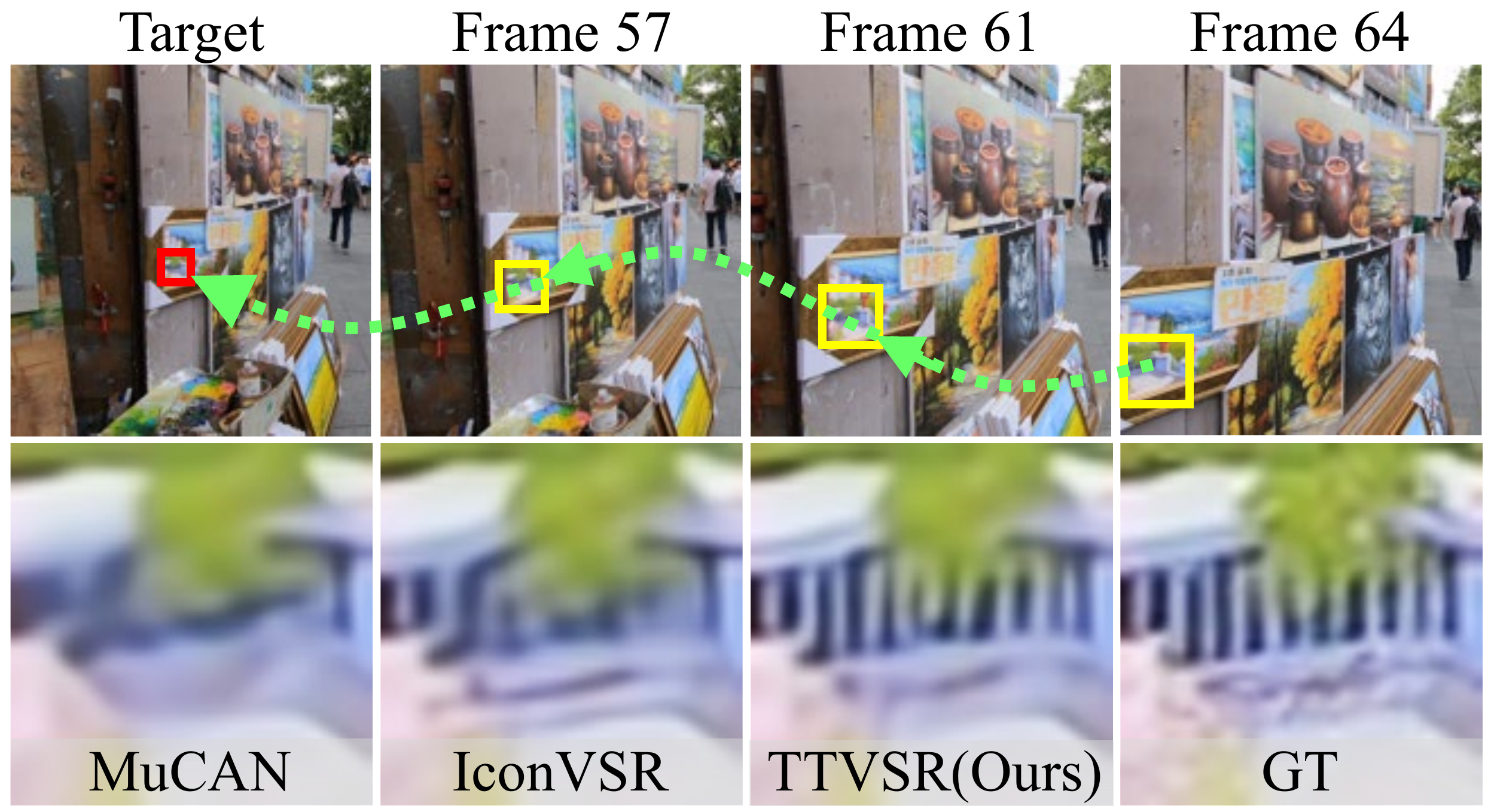}
  \vspace{-0.6cm}
  \caption{A comparison between TTVSR and other SOTA methods: MuCAN~\cite{li2020mucan} and IconVSR~\cite{chan2021basicvsr}. We introduce finer textures for recovering the target frame from the boxed areas (indicated by yellow) tracked by the trajectory (indicated by green).}
  \label{fig:teaser}
\end{figure}

Video super-resolution (VSR) aims to recover a high-resolution (HR) video from a low-resolution (LR) counterpart~\cite{wang2019edvr}. As a fundamental task in computer vision, VSR is usually adopted to enhance visual quality, which has great value in many practical applications, such as video surveillance~\cite{zhang2010super}, high-definition television~\cite{goto2014super}, and satellite imagery~\cite{luo2017video,deudon2020highres}, etc. From a methodology perspective, unlike image super-resolution that usually learns on spatial dimensions, VSR tasks pay more attention to exploiting temporal information. In Fig.~\ref{fig:teaser}, if detailed textures to recover the target frame can be discovered and leveraged at relatively distant frames, video qualities can be greatly enhanced.

To solve this challenge, recent years have witnessed an increasing number of VSR approaches, which can be categorized into two paradigms. The former makes attempts to utilize adjacent frames as inputs (e.g., 5 or 7 frames), and align temporal features in an implicit ~\cite{kim20183dsrnet,li2019fast} or explicit manners ~\cite{wang2019edvr,tian2020tdan}. One of the classic works is EDVR that adopts deformable convolutions to capture features within a sliding window~\cite{wang2019edvr}. However, larger window sizes will dramatically increase computational costs which makes this paradigm infeasible to capture distant frames. The latter investigates temporal utilization by recurrent mechanisms~\cite{sajjadi2018frame,yi2021omniscient,chan2021basicvsr}. One of the representative works is IconVSR that uses a hidden state to convey relevant features from entire video frames~\cite{chan2021basicvsr}. Nonetheless, recurrent networks usually lack long-term modeling capability due to vanishing gradient~\cite{hochreiter1998vanishing}, which inevitably leads to unsatisfied results as shown in Fig.~\ref{fig:teaser}.

Inspired by the recent progress of Transformer in natural language processing~\cite{vaswani2017attention}, significant progresses have been made in both visual recognition~\cite{carion2020end, dosovitskiy2020image} and generation tasks~\cite{yang2020learning, zeng2020learning}. For example, MuCAN proposes to use attention mechanisms to aggregate inter-frame features~\cite{li2020mucan} for VSR tasks. However, due to the high computational complexity in a video, it only learns from a narrow temporal window, which results in sub-optimal performance as shown in Fig.~\ref{fig:teaser}. Therefore, exploring proper ways of utilizing Transformers in videos remains a big challenge.

In this paper, we propose a novel Trajectory-aware Transformer to enable effective video representation learning for Video Super-Resolution (TTVSR). The key insight of TTVSR is to formulate video frames into pre-aligned trajectories of visual tokens, and calculate $\mathcal{Q}$, $\mathcal{K}$, and $\mathcal{V}$ in the same trajectory. In particular, we learn to link relevant visual tokens together along temporal dimensions, which forms multiple trajectories to depict object motions in a video (e.g., the green trajectory in Fig.~\ref{fig:teaser}). We update token trajectories by a proposed location map that online aggregates pixel motions around a token by average pooling. Once video trajectories have been learned, TTVSR calculates self-attention only on the most relevant visual tokens that are located in the same trajectory. Compared with MuCAN that calculates attention across visual tokens in space and time~\cite{li2020mucan}, the proposed TTVSR significantly reduces the computational cost and thus makes long-range video modeling practicable.

To further deal with the scale-changing problem that often occur in long-range videos (e.g., the yellow boxes in Fig.~\ref{fig:teaser}), we devise a cross-scale feature tokenization module and enhance feature representations from multiple scales. Our contributions are summarized as follows:
\begin{itemize}[nosep]
    \item We propose a novel trajectory-aware Transformer, which is one of the first works to introduce Transformer into video super-resolution tasks. Our method significantly reduces computational costs and enables long-range modeling in videos.
    \item Extensive experiments demonstrate that the proposed TTVSR can significantly outperform existing SOTA methods in four widely-used VSR benchmarks. In the most challenging REDS4 dataset, TTVSR gains 0.70db and 0.45db PSNR improvements than BasicVSR and IconVSR, respectively.
\end{itemize}

\section{Related Work}
\label{sec:related}

\subsection{Video Super-Resolution}
\label{sec:approach:video}
In VSR tasks, it is crucial to assist frame recovery with other frames in the sequence. Therefore, according to the number of input frames, VSR tasks can be mainly divided into two kinds of paradigms: based on sliding-window structure~\cite{caballero2017real,kim2018spatio,kim2019video,wang2019edvr,yi2019progressive,isobe2020video2,tian2020tdan,li2020mucan,xu2021temporal,cao2021video} and based on recurrent structure~\cite{huang2017video,sajjadi2018frame,fuoli2019efficient,haris2019recurrent,isobe2020revisiting,isobe2020video,yi2021omniscient,chan2021basicvsr}.

\noindent\textbf{Sliding-window structure.} The methods based on sliding-window structure use adjacent frames within a sliding window as inputs to recover the HR frame (e.g., 5 or 7 frames). They mainly focus on using 2D or 3D CNN~\cite{jo2018deep,isobe2020video2,li2019fast,kim20183dsrnet}, optical flow estimation~\cite{caballero2017real,tao2017detail,kim2018spatio} or deformable convolutions~\cite{wang2019edvr,tian2020tdan,dai2017deformable} to design advanced alignment modules and fuse detailed textures from adjacent frames. Typically, to fully utilize the complementary information across frames, FSTRN~\cite{li2019fast} presented a fast spatio-temporal residual network for VSR by adopting 3D convolutions~\cite{tran2015learning}. 
To better align adjacent frames, VESCPN~\cite{caballero2017real} introduced a spatio-temporal sub-pixel convolution network and first combined the motion compensation and VSR together. EDVR~\cite{wang2019edvr} and TDAN~\cite{tian2020tdan} used deformable convolutions~\cite{dai2017deformable} to align adjacent frames. However, they cannot utilize textures at other moments, especially in relatively distant frames.

\noindent\textbf{Recurrent structure.} Rather than aggregating information from adjacent frames, methods based on recurrent structure use a hidden state to convey relevant information in previous frames. FRVSR~\cite{sajjadi2018frame} used the previously SR frame to recover the subsequent frame. Inspired by the back-projection, RBPN~\cite{haris2019recurrent} treated each frame as a separate source, which is combined in an iterative refinement framework. RSDN~\cite{isobe2020video} divided the input into structure and detail components and proposed the two-steam structure-detail block to learn textures. Representatively, OVSR~\cite{yi2021omniscient}, BasicVSR~\cite{chan2021basicvsr}, and IconVSR~\cite{chan2021basicvsr} fused the bidirectional hidden state from the past and future for reconstruction and got significant improvements. They try to fully utilize the information of the whole sequence and synchronously update the hidden state by the weights of reconstruction network. However, due to the vanishing gradient~\cite{hochreiter1998vanishing}, this mechanism makes the updated hidden state loses the long-term modeling capabilities to some extent.

\begin{figure*}[t]
  \setlength{\belowcaptionskip}{-0.1cm}
  \centering
  \includegraphics[width=1.0\linewidth]{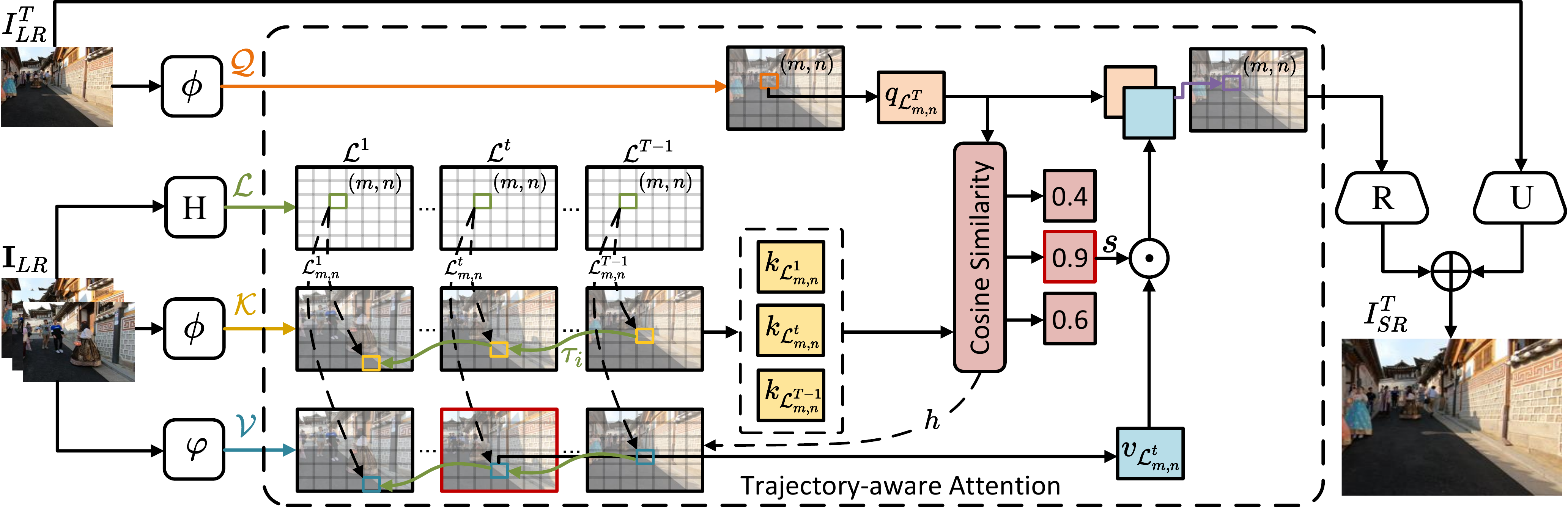}
  \vspace{-0.55cm}
\caption{The overview of TTVSR based on location maps. $\mathcal{Q}$, $\mathcal{K}$ and $\mathcal{V}$ are tokens from video frames extracted by embedding networks $\phi(\cdot)$ and $\varphi(\cdot)$, respectively. $\tau_{i}$ indicates a trajectory of $\mathcal{T}$. $\mathcal{L}$ is the set of location map generated by the motion estimation network $\text{H}$. The dotted lines indicates the indexing operation from $\mathcal{K}$ and $\mathcal{V}$ by location maps $\mathcal{L}$ and hard index $h$. $\text{R}(\cdot)$ represents the reconstruction network followed by a pixel-shuffle layer to resize feature maps to the desired size. $\text{U}(\cdot)$ represents the bicubic upsampling operation. $\odot$ and $\oplus$ indicate multiplication and element-wise addition, respectively.}
  \label{fig:overview}
    \vspace{-0.2cm}
\end{figure*}

\subsection{Vision Transformer}
\label{sec:approach:ViT}
Recently, Transformer~\cite{vaswani2017attention} has been proposed to improve the long-term modeling capabilities of sequence in various fields~\cite{devlin2018bert,dosovitskiy2020image}. In the field of computer vision~\cite{dosovitskiy2020image}, Transformer is used as a new attention-based module to model relationships between tokens in many image-based tasks, such as classification~\cite{dosovitskiy2020image}, inpainting~\cite{zeng2020learning}, super-resolution~\cite{yang2020learning}, 
generation~\cite{zeng2021improving} and so on. Typically, ViT~\cite{dosovitskiy2020image} unfolded an image into patches as tokens for attention to capture the long-range relationship in high-level vision. TTSR~\cite{yang2020learning} proposed a texture Transformer in low-level vision to search relevant texture patches from Ref image to LR image.

In VSR tasks, VSR-Transformer~\cite{cao2021video} and MuCAN~\cite{li2020mucan} tried to use attention mechanisms for aligning different frames with great success. However, due to the heavy computational costs of attention calculation on videos, these methods only aggregate information on the narrow temporal window. Therefore, in this paper, we introduce a trajectory-aware Transformer to improve the long-term modeling capabilities for VSR tasks while keeping the computational cost of attention within an acceptable range.

\section{Our Approach}
\label{sec:approach}
In this section, we first introduce the proposed \textbf{T}rajectory-aware \textbf{T}ransformer for \textbf{V}ideo \textbf{S}uper-\textbf{R}esolution (TTVSR) in Sec.~\ref{sec:approach:TT}, and then discuss the proposed location map for trajectory generation in Sec.~\ref{sec:approach:TG}. Finally, we refocus to our Transformer design based on the location maps and discuss its advantages in Sec.~\ref{sec:approach:LT}.

\subsection{Trajectory-Aware Transformer}
\label{sec:approach:TT}
We introduce the formulation of the TTVSR firstly, followed by trajectory-aware attention and cross-scale feature tokenization. More illustrations can be found in Fig.~\ref{fig:overview}.

\noindent\textbf{Formulation.}
Given a LR sequence, the goal of VSR tasks is to recover a HR version. Specifically, for our task, when restoring the $T^{\text{th}}$ frame $I_{SR}^{T}$, we denote the current LR frame as $I_{LR}^{T}$ and other LR frames as $\mathbf{I}_{LR}=\{I_{LR}^{t}, t \in [1,T-1]\}$. 

We use two embedding networks $\phi(\cdot)$ and $\varphi(\cdot)$ to get features from video frames and extract tokens by sliding-windows. The queries $\mathcal{Q}$ and keys $\mathcal{K}$ are extracted by $\phi(\cdot)$ and denoted as $\mathcal{Q}=\phi(I_{LR})=\{q_i^T,i\in [1,N]\}$ and $\mathcal{K}=\phi(\mathbf{I}_{LR})=\{k_i^t,i\in [1,N],t\in [1, T-1]\}$, respectively. The values are extracted by $\varphi(\cdot)$ and denoted as $\mathcal{V}=\varphi(\mathbf{I}_{LR})=\{v_i^t,i\in [1,N],t\in [1, T-1]\}$.

The trajectories $\mathcal{T}$ in our approach can be formulated as a set of trajectory, in which each trajectory $\tau_i$ is a sequence of coordinate over time and the end point of trajectory $\tau_i$ is associated with the coordinate of token $q_i$:

\begin{equation}
\begin{aligned}
\mathcal{T} &=\{\tau_i, \:i\in [1,N]\},\\
\tau_i &=\langle \tau_i^t=(x_i^t, y_i^t),\: t\in [1,T] \rangle,\\
\end{aligned}
\label{equ:deftrj}
\end{equation}
where $x_i^t\in [1,H]$, $y_i^t\in [1,W]$, and $(x_i^t, y_i^t)$ represents the coordinate of trajectory $\tau_i$ at time $t$. $H$ and $W$ represents the height and width of the feature maps, respectively.

From the aspect of trajectories, the inputs of proposed trajectory-aware transformer can be further represented as visual tokens which are aligned by trajectories $\mathcal{T}$:
\begin{align}
\begin{aligned}
&\mathcal{T}=\{\tau_i, \:i\in [1,N]\},\\
&\mathcal{Q}=\{q_{\tau_i^T},\:i\in [1,N]\},\\
&\mathcal{K}=\{k_{\tau_i^t},\:i\in [1,N],\:t\in [1, T-1]\},\\
&\mathcal{V}=\{v_{\tau_i^t},\:i\in [1,N],\:t\in [1, T-1]\}.
\end{aligned}
\end{align}

The process of recovering the $T^{\text{th}}$ HR frame $I_{SR}^T$ can be further expressed as:
\begin{equation}
\begin{aligned}
I_{SR}^T&=\text{T}_{traj}(\mathcal{Q},\mathcal{K},\mathcal{V},\mathcal{T})\\
 &=\text{R}(\mathop{\text{A}_{traj}}_{\tau_i\in \mathcal{T}}(q_{\tau_i^T},k_{\tau_i^t},v_{\tau_i^t}))+\text{U}(I_{LR}^T),
\end{aligned}
\label{equ:ttvsr}
\end{equation}
where $\text{T}_{traj}(\cdot)$ denotes the trajectory-aware Transformer. $\text{A}_{traj}(\cdot)$ denotes the trajectory-aware attention. $\text{R}(\cdot)$ represents the reconstruction network followed by a pixel-shuffle layer to resize feature maps to the desired size. $\text{U}(\cdot)$ represents the bicubic upsampling operation.

By introducing trajectories into Transformer, the attention calculation on $\mathcal{K}$ and $\mathcal{V}$ can be significantly reduced because it can avoid the computation on spatial dimension compared with vanilla vision Transformers.

\noindent\textbf{Trajectory-aware attention.}
Thanks to the powerful long-range model ability, the attention mechanisms in vanilla vision Transformer is used to model dependencies of tokens within an image~\cite{dosovitskiy2020image,carion2020end}. However, empowering the attention mechanisms to videos remains a challenge.
Thus, we propose a trajectory-aware attention module, which integrates relevant visual tokens located in the same spatio-temporal trajectories with less computational costs.

Different from the traditional attention mechanisms that take a weighted sum of keys in temporal. We use hard attention to select the most relevant token along trajectories, its purpose is to reduce blur introduced by weighted sum. We use soft attention to generate the confidence of relevant patches, it is used to reduce the impact of irrelevant tokens when hard attention gets inaccurate results. We use $h_{\tau_i}$ and $s_{\tau_i}$ to represent the results of hard and soft attention. The calculation process can be formulated as:
\begin{equation}
\begin{split}
h_{\tau_i} & = \mathop{\arg\max}\limits_{t}{\langle \frac{q_{\tau_i^T}}{{\parallel q_{\tau_i^T} \parallel}_{2}^{2}}, \frac{k_{\tau_i^t}}{{\parallel k_{\tau_i^t} \parallel}_{2}^{2}} \rangle}, \\
s_{\tau_i} & = \mathop{\max}\limits_{t}{\langle \frac{q_{\tau_i^T}}{{\parallel q_{\tau_i^T} \parallel}_{2}^{2}}, \frac{k_{\tau_i^t}}{{\parallel k_{\tau_i^t} \parallel}_{2}^{2}} \rangle}.
\end{split}
\end{equation}
Based on such formula, the attention calculation in Equ.~\ref{equ:ttvsr} can be formulated as:
\begin{equation}
\text{A}_{traj}(q_{\tau_i^T},k_{\tau_i},v_{\tau_i}) = \text{C}(q_{\tau_i^T}\:,\:{s_{\tau_i} \odot v_{\tau_i^{h_{\tau_i}}}}),
\end{equation}
where the operator $\odot$ denotes multiplication. $\text{C}(\cdot)$ denotes the concatenation operation. We fold all the tokens and output a feature map.

In general, in the proposed trajectory-aware attention, we integrate features from the whole sequence. Such a design allows attention calculation only along its spatio-temporal trajectory, mitigating the computational cost.

\noindent\textbf{Cross-scale feature tokenization.}
\label{sec:approach:CFT}
The premise of utilizing multi-scale texture from sequences is that the model can adapt to the multi-scale variations in content that often occur.
Therefore, we propose a cross-scale feature tokenization module before trajectory-aware attention to extract tokens from multiple scales. It can uniform multi-scale features into a uniform-length token and allows rich textures from larger scales to be utilized for the recovery of smaller ones in the attention mechanism.

Specifically, we follow three steps to extract tokens.
First, the successive unfold and fold operations are used to expand the receptive field of features.
Second, features from different scales are shrunk to the same scale by a pooling operation.
Third, the features are split by unfolding operation to obtain the output tokens.
It is noteworthy that this process can extract features from a larger scale while keeping the same size as output tokens. It is convenient for attention calculation and token integration. More analyses can be found in the supplementary.

\begin{figure}[t]
  \setlength{\belowcaptionskip}{-0.1cm}
  \centering
  \includegraphics[width=1.0\linewidth]{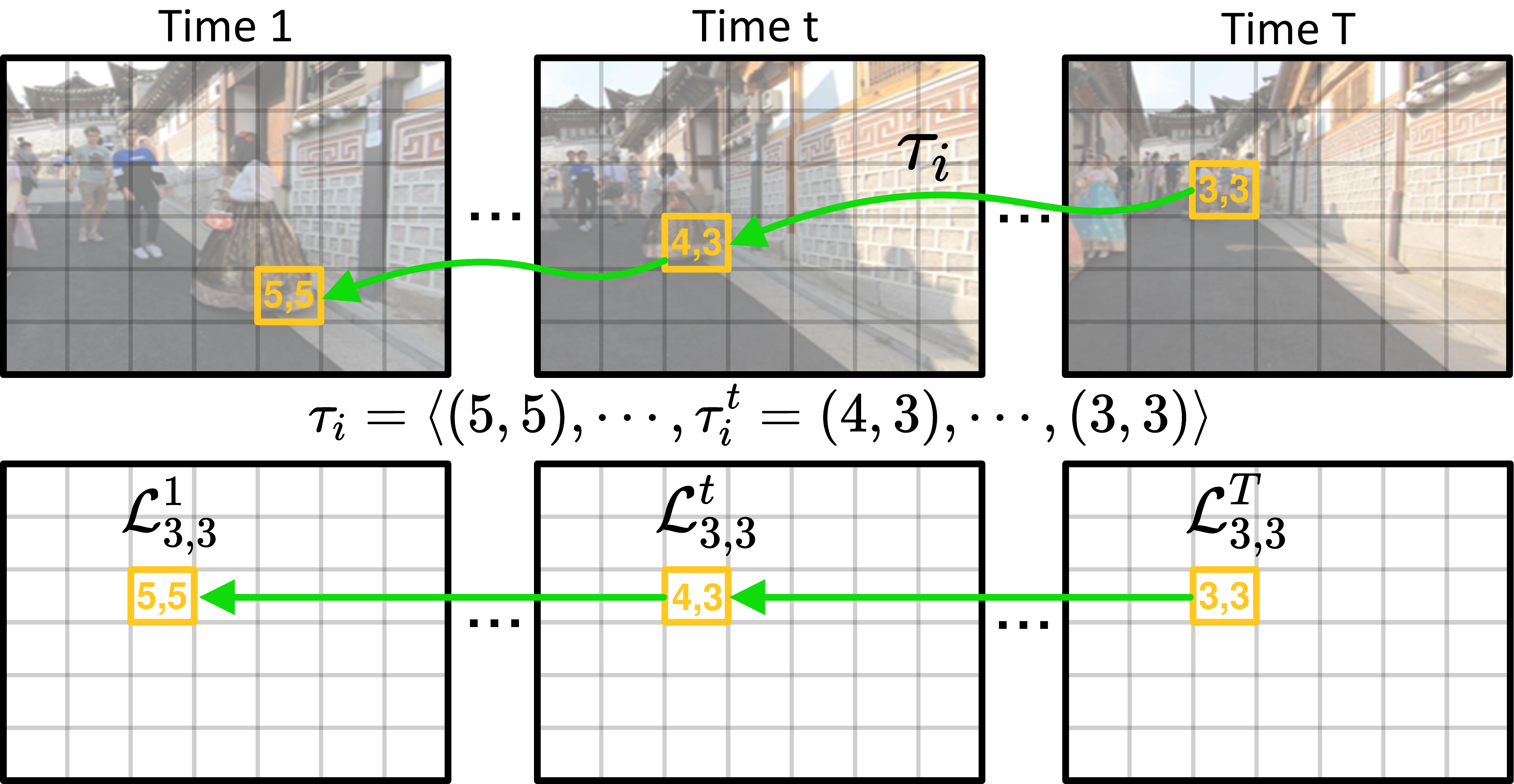}
  \vspace{-0.6cm}
  \caption{An illustration of the relationship between trajectory $\tau$ and the location maps $\mathcal{L}$ at time $t$.}
  \label{fig:locmap}
    \vspace{-0.2cm}
\end{figure}

\subsection{Location Maps for Trajectory Generation}
\label{sec:approach:TG}
Existing approaches use feature alignment and global optimization to calculate trajectories of video which are time-consuming and less efficient~\cite{wang2013dense,wang2013action,patrick2021keeping}. Especially in our task, trajectories are updated over time, the computation cost will be further exploded. To solve this problem, we propose a location map for trajectory generation in which the location maps are represented as a group of matrices over time. By such a design, the trajectory generation can be expressed as some matrix operations which are both efficient for computing and friendly for model implementation.

Since the trajectories are updated over time, our location maps need also to be updated accordingly. In the formulation of it, we fix the time to $T$ for better illustration. The proposed location maps can be formulated as:
\begin{equation}
\mathcal{L}^{t} = \begin{bmatrix}
      (x_1,y_1) & \dots & (x_1,y_W) \\
    \dots &  \dots &  \dots \\
      (x_H,y_1) &  \dots & (x_H,y_W)\\
      \end{bmatrix},\:t\in [1,T],
\end{equation}
where $\mathcal{L}_{m,n}^{t}$ represents the coordinate at time $t$ in a trajectory which is ended at $(m, n)$ at time $T$. The relationship between the location map $\mathcal{L}^{t}_{m,n}$ and the trajectory $\tau_i^t$ defined in Equ.~\ref{equ:deftrj} can be further expressed as:
\begin{equation}
\mathcal{L}_{m,n}^{t}=\tau_i^t,\: \text{where}\ \tau_i^T=(m,n),\: i\in [1,N],
\label{equ:lmdef}
\end{equation}
where $m\in [1,H]$ and $n\in [1,W]$. In Fig.~\ref{fig:locmap}, we use a simple case to further illustrate the relationship between location maps and trajectories.

\noindent\textbf{Location map updating.} As discussed in the formulation part, the location maps will change over time. We denotes the updated location maps as ${}^*\!\mathcal{L}^{t}$. When changing from time $T$ to time $T+1$, a new location map ${}^*\!\mathcal{L}^{T+1}$ at time $T+1$ should be initialized. Based on Equ.~\ref{equ:lmdef}, the element values of ${}^*\!\mathcal{L}^{T+1}$ are exactly the coordinates of frame $T+1$~\footnote{Where the element values of the matrix are equal to the index matrix.}.

Then the rest updated location maps $\{{}^*\!\mathcal{L}^{1},\cdots,{}^*\!\mathcal{L}^{T}\}$ can be obtained by tracking the location maps $\{\mathcal{L}^{1},\cdots,\mathcal{L}^{T}\}$ from time $T+1$ to time $T$ using backward flow $O^{T+1}$. Specifically, $O^{T+1}$ can build the connection of trajectories between time $T$ and time $T+1$ and obtain from a lightweight motion estimation network. Due to the correlations in flow are usually float numbers, we get the updated coordinates in location map $\mathcal{L}^{t}$ by interpolating between its adjacent coordinates:
\begin{equation}
{}^*\!\mathcal{L}^{t} = \text{S}(\mathcal{L}^{t}, O^{T+1}), 
\label{motion}
\end{equation}
where $\text{S}(\cdot)$ represents the spatial sampling operation by spatial correlation $O^{T+1}$ (i.e., $grid\_sample$ in PyTorch). Thus far, we have all the updated location maps for time $T+1$.

With the careful design of the location maps, the trajectories in our proposed trajectory-aware Transformer can be effectively calculated and maintained through one parallel matrix operation (i.e., the operation $\text{S}(\cdot)$). More analyses can be found in the supplementary.

\subsection{TTVSR based on Location Maps}
\label{sec:approach:LT}
In this section, we recap the formulation of our proposed TTVSR in Sec.~\ref{sec:approach:TT} and show the relation between TTVSR and location maps in a more intrinsical way. More details can be found in Fig.~\ref{fig:overview}. Since the location map $\mathcal{L}^{t}$ in Equ.~\ref{equ:lmdef} is an interchangeable formulation of trajectory $\tau_i$ in Equ.~\ref{equ:ttvsr}, the proposed TTVSR can be further expressed as:
\begin{align}
\begin{aligned}
I_{SR}^T&=\text{T}_{traj}(\mathcal{Q},\mathcal{K},\mathcal{V},\mathcal{L})\\
 &=\text{R}(\mathop{\text{A}_{traj}}_{t,m,n}(q_{\mathcal{L}_{m,n}^{T}},k_{\mathcal{L}_{m,n}^{t}},v_{\mathcal{L}_{m,n}^{t}}))+\text{U}(I_{LR}^T),
\end{aligned}
\end{align}
where $m\in [1, H]$, $n\in [1, W]$, and $t\in [1,T-1]$.

In this formulation, we transform the coordinate system in our transformer from the one defined by trajectories to a group of aligned matrices (i.e., the location maps). Such a design has two advantages: First, the location maps provide a more efficient way to enable our TTVSR can directly leverage the information from a distant video frame. Second, as the trajectory is a widely used concept in videos, our design can motivate other video tasks to achieve a more efficient and powerful implementation.

\subsection{Training Details}
\label{sec:approach:patch}

\par
    For fair comparisons, we follow IconVSR~\cite{chan2021basicvsr} and VSR-Transformer~\cite{cao2021video} to use the same feature extraction network, reconstruction network, and pre-trained SPyNet~\cite{ranjan2017optical} for motion estimation. To leverage the information of the whole sequence, we follow previous works~\cite{huang2017video,chan2021basicvsr} to adopt a bidirectional propagation scheme, where the features in different frames can be propagated backward and forward, respectively. To reduce consumption in terms of time and memory, we generate the visual tokens of different scales from different frames. 
    Features from adjacent frames are finer, so we generate tokens of size $1 \times1$. Features from a long distance are coarser, so we select these frames at a certain time interval and generate tokens of size $4 \times4$. Besides, in Sec.~\ref{sec:approach:CFT}, we use kernels of size $4 \times4$, $6 \times6$, and $8 \times8$ for cross-scale feature tokenization. During training, we use Cosine Annealing scheme~\cite{loshchilov2016sgdr} and Adam~\cite{kingma2014adam} optimizer with $\beta_{1}=0.9$ and $\beta_{2}=0.99$. The learning rates of the motion estimation and other parts are set as $1.25\times 10^{-5}$ and $2\times 10^{-4}$, respectively. We set the batch size as $8$ and input patch size as $64\times 64$. To keep fair comparison, we augment the training data with random horizontal flips, vertical flips, and $90^{\circ}$ rotations. Besides, to enable long-range sequence capability, we use sequences with a length of 50 as inputs. The Charbonnier penalty loss~\cite{lai2017deep} is applied on whole frames between the ground-truth $I_{HR}$ and restored SR frame $I_{SR}$, which can be defined by $\ell=\sqrt{\|I_{HR}-I_{SR}\|^{2}+\varepsilon^2}$. To stabilize the training of TTVSR, we fix the weights of the motion estimation module in the first 5K iterations, and make it trainable later. The total number of iterations is 400K.

\section{Experiments}
\label{sec:experiments}

\setlength{\tabcolsep}{1.0mm}{
\begin{table*}\small
  \caption{Quantitative comparison (PSNR$\uparrow$ and SSIM$\uparrow$) on the REDS4~\cite{nah2019ntire} dataset for $4\times$ video super-resolution. The results are tested on RGB channels. \textcolor{red}{Red} indicates the best and \textcolor{blue}{blue} indicates the second best performance (best view in color). \#Frame indicates the number of input frames required to perform an inference, and ``r'' indicates to adopt the recurrent structure. 
  }
  \vspace{-0.2cm}
  \centering
  \begin{tabular}{ l  || c || c | c | c | c || c }
    \hline
    Method  &  \#Frame     &  Clip\_000 &  Clip\_011 &  Clip\_015 &  Clip\_020  &  Average  \\
    \hline
    Bicubic                     &       1    &   24.55/0.6489  &  26.06/0.7261  &   28.52/0.8034    &   25.41/0.7386   &  26.14/0.7292 \\
    RCAN~\cite{zhang2018image}   &       1    &   26.17/0.7371   &  29.34/0.8255 &   31.85/0.8881  &   27.74/0.8293   &   28.78/0.8200  \\
    CSNLN~\cite{mei2020image}    &       1    &    26.17/0.7379  &   29.46/0.8260    &  32.00/0.8890   &   27.69/0.8253   &  28.83/0.8196  \\
    \hline
    TOFlow~\cite{xue2019video}    &       7  &    26.52/0.7540  &   27.80/0.7858   &  30.67/0.8609    &  26.92/0.7953    &   27.98/0.7990 \\
    DUF~\cite{jo2018deep}         &   7   &   27.30/0.7937  &  28.38/0.8056   &  31.55/0.8846  &   27.30/0.8164  &  28.63/0.8251  \\
    EDVR~\cite{wang2019edvr}      &   7  &  28.01/0.8250  &  32.17/0.8864  &  34.06/0.9206  &   30.09/0.8881   &   31.09/0.8800\\
    MuCAN~\cite{li2020mucan}      &       5   &   27.99/0.8219  &  31.84/0.8801   &   33.90/0.9170  & 29.78/0.8811    &  30.88/0.8750  \\
    VSR-T~\cite{cao2021video}     &     5   &   28.06/0.8267  &  32.28/0.8883  &  34.15/0.9199  &  30.26/0.8912  &   31.19/0.8815 \\
    \hline
    BasicVSR~\cite{chan2021basicvsr}& r & 28.39/0.8429 & 32.46/0.8975 & 34.22/0.9237 & 30.60/0.8996 & 31.42/0.8909 \\
    IconVSR~\cite{chan2021basicvsr}&  r   &\textcolor{blue}{28.55}/\textcolor{blue}{0.8478}& \textcolor{blue}{32.89}/\textcolor{blue}{0.9024} & \textcolor{blue}{34.54}/\textcolor{blue}{0.9270} & \textcolor{blue}{30.80}/\textcolor{blue}{0.9033} &  \textcolor{blue}{31.67}/\textcolor{blue}{0.8948} \\
    \hline
    \textbf{TTVSR}            &  r  & \textcolor{red}{28.82}/\textcolor{red}{0.8566}& \textcolor{red}{33.47}/\textcolor{red}{0.9100} & \textcolor{red}{35.01}/\textcolor{red}{0.9325} & \textcolor{red}{31.17}/\textcolor{red}{0.9094} & \textcolor{red}{32.12}/\textcolor{red}{0.9021} \\
    \hline    
  \end{tabular}
  \label{tab:BI}
\vspace{-0.25cm}
\end{table*}
}

\setlength{\tabcolsep}{1.0mm}{
\begin{table}\small
  \caption{Quantitative comparison (PSNR$\uparrow$ and SSIM$\uparrow$) on Vid4~\cite{liu2013bayesian}, UDM10~\cite{yi2019progressive} and Vimeo-90K-T~\cite{xue2019video} dataset for $4\times$ video super-resolution. All the results are calculated on Y-channel. \textcolor{red}{Red} indicates the best and \textcolor{blue}{blue} indicates the second best performance (best view in color).}
  \vspace{-0.2cm}
  \centering
  \begin{tabular}{ l ||  c | c | c }
    \hline
    Method & Vid4~\cite{liu2013bayesian} & UDM10~\cite{yi2019progressive} & Vimeo-90K-T~\cite{xue2019video} \\
    \hline
    Bicubic         &    21.80/0.5246    &     28.47/0.8253&    31.30/0.8687 \\
    TOFlow~\cite{xue2019video}   &   25.85/0.7659     &    36.26/0.9438 &    34.62/0.9212    \\
    FRVSR~\cite{sajjadi2018frame}     &   26.69/0.8103    &     37.09/0.9522   &    35.64/0.9319 \\
    DUF~\cite{jo2018deep}    &    27.38/0.8329    &    38.48/0.9605  &    36.87/0.9447   \\
    RBPN~\cite{haris2019recurrent}   &    27.17/0.8205    &     38.66/0.9596  &    37.20/0.9458   \\
    RLSP~\cite{fuoli2019efficient} &    27.48/0.8388    &     38.48/0.9606  &   36.49/0.9403    \\
    EDVR~\cite{wang2019edvr}     &    27.85/0.8503     &    39.89/0.9686  &    37.81/0.9523   \\
    TDAN~\cite{tian2020tdan}     &   26.86/0.8140    &    38.19/0.9586 &    36.31/0.9376      \\
    TGA~\cite{isobe2020video2}    &    27.59/0.8419    &     39.05/0.9634     &   37.59/0.9516       \\
    RSDN~\cite{isobe2020video}        &    27.92/0.8505   &     39.35/0.9653  &    37.23/0.9471  \\
    BasicVSR~\cite{chan2021basicvsr} &    27.96/0.8553    &     39.96/0.9694 &    37.53/0.9498   \\
    IconVSR~\cite{chan2021basicvsr} &   \textcolor{blue}{28.04}/\textcolor{blue}{0.8570}    &     \textcolor{blue}{40.03}/\textcolor{blue}{0.9694} &   \textcolor{blue}{37.84}/\textcolor{blue}{0.9524}   \\
    \hline
    \textbf{TTVSR}  &  \textcolor{red}{28.40}/\textcolor{red}{0.8643} &     \textcolor{red}{40.41}/\textcolor{red}{0.9712}  &  \textcolor{red}{37.92}/\textcolor{red}{0.9526} \\  
    \hline    
  \end{tabular}
  \label{tab:BD}
\vspace{-0.3cm}
\end{table}
}

\subsection{Datasets and Metrics}
We evaluate the proposed TTVSR and compare its performance with other SOTA approaches on two widely-used datasets: \textbf{REDS}~\cite{nah2019ntire} and \textbf{Vimeo-90K}~\cite{xue2019video}. For \textbf{REDS}~\cite{nah2019ntire}, it is published in the NTIRE19 challenge~\cite{nah2019ntire}. It contains a total of 300 video sequences, in which 240 for training, 30 for validation, and 30 for testing. Each sequence contains 100 frames with a resolution of $720 \times 1280$. 
To create training and testing sets, we follow previous works~\cite{wang2019edvr,li2020mucan,chan2021basicvsr} to select four sequences\footnote{Clips 000,011,015,020 of the REDS training set.} as the testing set which is called \textbf{REDS4}~\cite{nah2019ntire}. And we select the rest 266 sequences from the training and validation set as the training set.
For \textbf{Vimeo-90K}~\cite{xue2019video}, it contains 64,612 sequences for training and 7,824 for testing. Each sequence contains seven frames with a resolution of $448 \times 256$. For fair comparison, we follow previous works~\cite{chan2021basicvsr} to evaluate TTVSR with $4 \times$ downsampling by using two degradations: 1) MATLAB bicubic downsample (BI), and 2) Gaussian filter with a standard deviation of $\sigma=1.6$ and downsampling (BD). Same with previous works~\cite{isobe2020video,isobe2020video2,tian2020tdan,li2020mucan}, we apply the BI degradation on \textbf{REDS4}~\cite{nah2019ntire} and BD degradation on \textbf{Vimeo-90K-T}~\cite{xue2019video}, \textbf{Vid4}~\cite{liu2013bayesian} and \textbf{UDM10}~\cite{yi2019progressive}. 
We keep the same evaluation metrics: 1) Peak signal-to-noise ratio (PSNR) and 2) structural similarity index (SSIM)~\cite{wang2004image} as previous works \cite{li2020mucan,chan2021basicvsr}.

\subsection{Comparisons with State-of-the-art Methods}
\par
We compare TTVSR with 15 start-of-the-art methods. These methods can be summarized into three categories: single image super-resolution (SISR)~\cite{zhang2018image,mei2020image}, sliding window-based~\cite{xue2019video,jo2018deep,wang2019edvr,tian2020tdan,isobe2020video2,li2020mucan,cao2021video}, and recurrent structure-based ~\cite{sajjadi2018frame,fuoli2019efficient,haris2019recurrent,isobe2020video,chan2021basicvsr}.
For fair comparisons, we obtain the performance from their original paper or reproduce results by authors officially released models.

\par
\noindent\textbf{Quantitative comparison.} We compare TTVSR with other SOTA methods on the most widely-used REDS dataset~\cite{nah2019ntire}. As shown in Tab.~\ref{tab:BI}, we categorize these approaches according to the frames used in each inference. Among them, since only one LR frame is used, the performance of SISR methods~\cite{zhang2018image,mei2020image} is very limited. MuCAN~\cite{li2020mucan} and VSR-T~\cite{cao2021video} use attention mechanisms in sliding window, which has a significant improvement over the SISR methods. However, they do not fully utilize the information of the sequence. BasicVSR~\cite{chan2021basicvsr} and IconVSR~\cite{chan2021basicvsr} try to model the whole sequence through hidden states. 
Nonetheless, the well-known vanishing gradient issue limits their capabilities of long-term modeling, thus the information at a distance will be lost. 
Different from them, our TTVSR tries to link the relevant visual token together along the same trajectory in an efficient way. TTVSR also uses the whole sequence information to recover the lost textures. Due to such merits, TTVSR achieves a result of 32.12dB PSNR and significantly outperforms IconVSR~\cite{chan2021basicvsr} by \textbf{0.45dB} on the REDS4~\cite{nah2019ntire}. This large margin demonstrates the power of TTVSR in long-range modeling.

\setlength{\tabcolsep}{1.0mm}{
\begin{table}\small
  \caption{Comparison of params, FLOPs and numbers. FLOPs is computed on one LR frame with the size of $180 \times320$ and $\times 4$ upsampling on the REDS4~\cite{nah2019ntire} dataset.}
  \vspace{-0.2cm}
  \centering
  \begin{tabular}{ l || c || c || c}
    \hline
    Method            &        \#Params(M)          &       FLOPs(T)           &      PSNR/SSIM    \\
    \hline
    DUF\cite{jo2018deep}            &        5.8       &       2.34           &       28.63/0.8251      \\
    RBPN\cite{haris2019recurrent}   &        12.2      &       8.51          &       30.09/0.8590      \\
    EDVR\cite{wang2019edvr}         &        20.6      &       2.95         &       31.09/0.8800      \\
    MuCAN\cite{li2020mucan}        &         13.6      &       $>$1.07        &       30.88/0.8750      \\
    BasicVSR\cite{chan2021basicvsr} &        6.3       &       0.33          &       31.42/0.8909      \\
    IconVSR\cite{chan2021basicvsr}  &        8.7       &       0.51         &       31.67/0.8948      \\
    \textbf{TTVSR}             &        6.8       &       0.61        &       32.12/0.9021      \\
    \hline    
  \end{tabular}
  \label{tab:MS}
  \vspace{-0.3cm}
\end{table}
}

\begin{figure*}[t!]
\setlength{\belowcaptionskip}{-0.2cm}
\setlength{\abovecaptionskip}{0.2cm}
  \centering
  \includegraphics[width=1.0\linewidth]{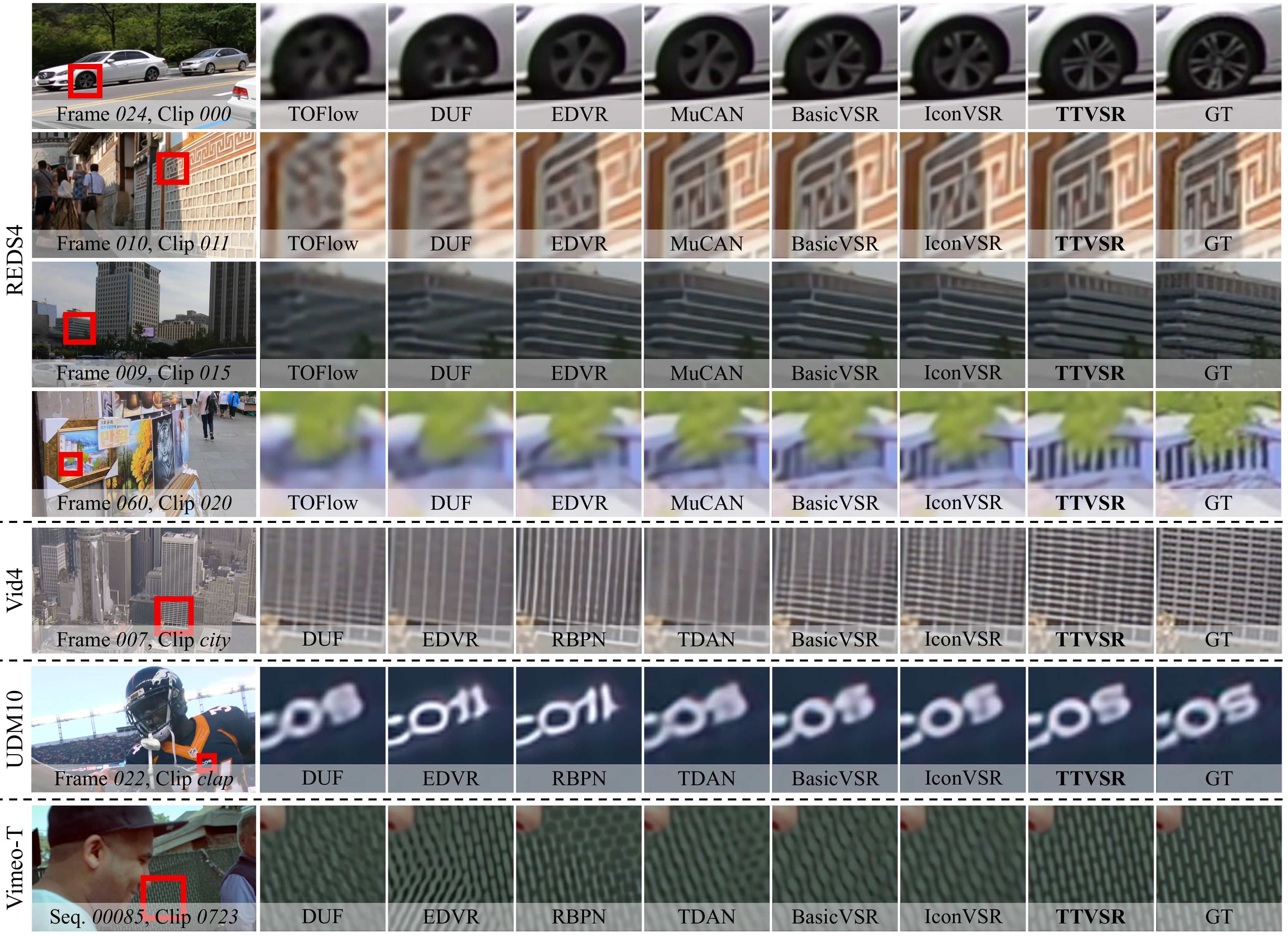}
  \vspace{-0.55cm}
   \caption{Visual results on REDS4~\cite{nah2019ntire}, Vid4~\cite{liu2013bayesian}, UDM10~\cite{yi2019progressive} and Vimeo-90K-T~\cite{xue2019video} for $4 \times$ scaling factor. The frame number is shown at the bottom of each case. Zoom in to see better visualization.}
   \label{fig:case}
   \vspace{-0.25cm}
\end{figure*}

To further verify the generalization capabilities of TTVSR, we train TTVSR on Vimeo-90K dataset~\cite{xue2019video}, and evaluate the results on Vid4~\cite{liu2013bayesian}, UDM10~\cite{yi2019progressive}, and Vimeo-90K-T datasets~\cite{xue2019video}, respectively. As shown in Tab.~\ref{tab:BD}, on the Vid4~\cite{liu2013bayesian}, UDM10~\cite{yi2019progressive}, and Vimeo-90K-T~\cite{xue2019video} test sets, TTVSR achieves the results of 28.40dB, 40.41dB, and 37.92dB in PSNR respectively, which is superior to other SOTA methods. Specifically, on the Vid4~\cite{liu2013bayesian} and UDM10~\cite{yi2019progressive} datasets, TTVSR outperforms IconVSR~\cite{chan2021basicvsr} by \textbf{0.36dB} and \textbf{0.38dB} respectively. At the same time, we notice that compared with the evaluation on Vimeo-90K-T~\cite{xue2019video} dataset with only seven frames in each testing sequence, TTVSR has a better improvement on other datasets which have at least 30 frames per video. The results verify that TTVSR has strong generalization capabilities and is good at modeling the information in long-range sequences.

\noindent\textbf{Qualitative comparison.} To further compare visual qualities of different approaches, we show visual results generated by TTVSR and other SOTA methods on four different test sets in Fig.~\ref{fig:case}. For fair comparisons, we either directly take the original SR images of the author-released or use author-released models to get results. It can be observed that TTVSR has a great improvement in visual quality, especially for areas with detailed textures. For example, in the fourth row in Fig.~\ref{fig:case}, TTVSR can recover more striped details from the stonework in the oil painting. The results verify that TTVSR can utilize textures from relevant tokens to produce finer results. More visual results can be found in the supplementary materials.

\par
\noindent\textbf{Model sizes and computational costs.} In real applications, model sizes and computational costs are usually important. To avoid the gap between different hardware devices, we use two hardware-independent metrics, including the number of parameters (\#Params) and FLOPs. As shown in Tab.~\ref{tab:MS}, the FLOPs are computed with the input of LR size $180 \times320$ and $\times 4$ upsampling settings. Compared with IconVSR~\cite{chan2021basicvsr}, TTVSR achieves higher performance while keeping comparable \#Params and FLOPs. Besides, it should be emphasized that our method is much lighter than MuCAN~\cite{li2020mucan} which is the SOTA attention-based method. Such superior performances mainly benefit from the use of trajectories in attention calculation which significantly reduces  computational costs.

\setlength{\tabcolsep}{1.0mm}{
\begin{table}\small
  \caption{Ablation study results of trajectory-aware attention module on the REDS4~\cite{nah2019ntire} dataset. TG: trajectory generation. TA: trajectory-aware attention.}
  \centering
  \vspace{-0.2cm}
  \begin{tabular}{ l  || c | c || c}
    \hline
   Method           &        TG          &      TA      &      PSNR/SSIM    \\
    \hline
    Base              &        ~           &          ~         &    30.46/0.8661   \\
    Base+TG        &        $\checkmark$        &           ~            &    31.91/0.8985   \\
    Base+TG+TA     &        $\checkmark$            &       $\checkmark$      &     \textbf{31.99}/\textbf{0.9007}   \\
    \hline    
  \end{tabular}
  \label{tab:TA}
\vspace{-0.3cm}
\end{table}}

\subsection{Ablation Study}
In this section, we conduct the ablation study on the proposed trajectory-aware attention and study the influence of frames number used in this module. In addition, we further analyze the effect of the cross-scale feature tokenization. 

\begin{figure}
\setlength{\belowcaptionskip}{-0.2cm}
\setlength{\abovecaptionskip}{0.1cm}
  \centering
  \includegraphics[width=1.0\linewidth]{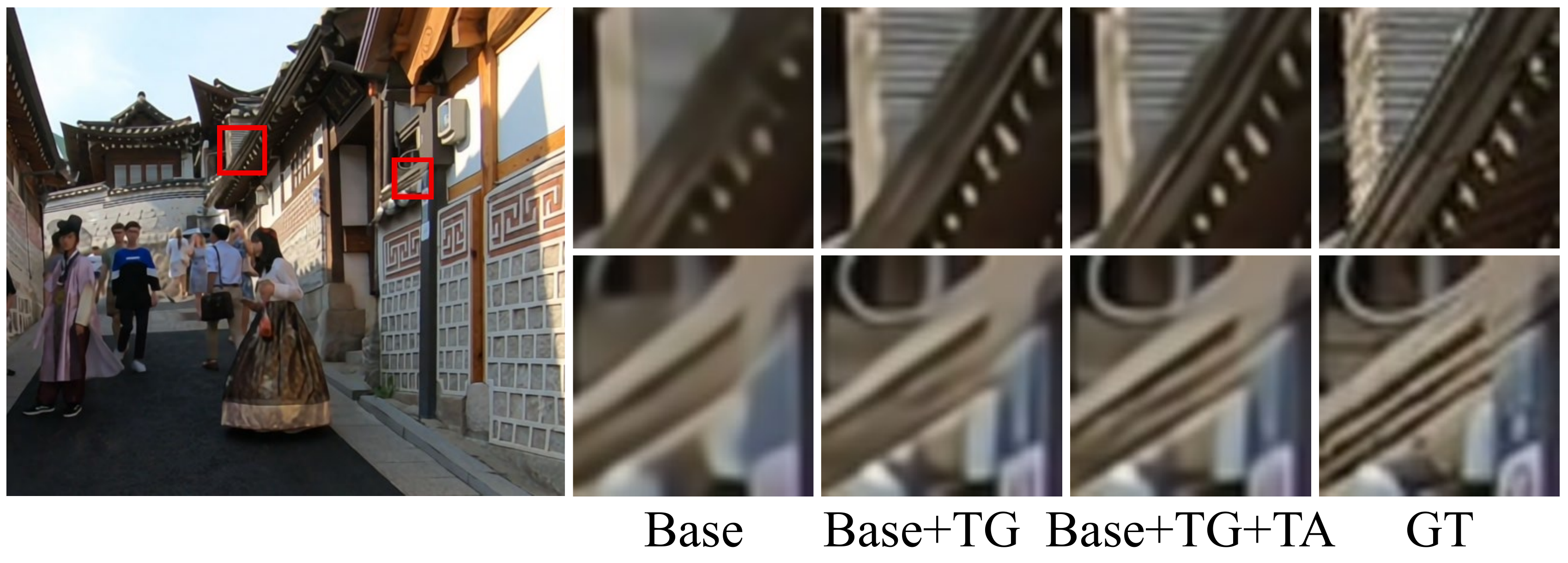}
  \vspace{-0.5cm}
  \caption{Ablation study on the trajectory generation (TG) and trajectory-aware attention (TA) on the REDS4~\cite{nah2019ntire} dataset.}
  \label{fig:ab_ta}
  \vspace{-0.3cm}
\end{figure}

\noindent\textbf{Trajectory-aware attention.} Trajectory generation (TG) is a prerequisite for trajectory-aware attention (TA), so we study them together in this part. We directly use convolution layers to integrate the not aligned previous tokens and current token as our ``Base" model. We denote the model that aggregates the most relevant tokens on the trajectory as our ``Base+TG" model. We denote the model that adds trajectory-aware attention progressively as our ``Base+TG+TA" model. The results are shown in Tab.~\ref{tab:TA}. With the addition of TG, PSNR can be improved from 30.46 to 31.91, which verifies that the trajectory can link relevant visual tokens together precisely. When TA is involved, we integrate tokens from trajectories, and the performance is improved to 31.99. This demonstrates the superiority of TA for modeling long-range information. We further explore the visual differences as shown in Fig.~\ref{fig:ab_ta}. TG can capture the relevant tokens, while TA integrates tokens into the current frame to produce clearer textures. 

\noindent\textbf{Influence of frame number during inference.} To explore the influence of the frame number used during inference on the ability of modeling long-range sequences. As shown in Tab.~\ref{tab:ab_lf}, we use different temporal intervals to sample frames from the entire sequence (100 frames). The performance is positively correlated with the number of sampled frames. It demonstrates the effectiveness of the trajectory-aware attention module for long-range modeling. However, the performance gain gradually decreases when the frame number is more than 45. It indicates that choosing three as the temporal interval (i.e., 33 frames) is sufficient to model the entire sequences. Using smaller intervals may not provide more information since the adjacent frames are too similar.

\setlength{\tabcolsep}{1.0mm}{
\begin{table}\small
  \caption{Ablation study results of the frame number used on the REDS4~\cite{nah2019ntire} dataset.}
  \centering
  \vspace{-0.2cm}
  \begin{tabular}{ l || c | c | c | c | c}
    \hline
    \#Frame  &  5  &  10  &   20  &  33 &  45\\
    \hline
    PSNR  &31.89 & 31.93 & 31.97 & 31.99 & 32.01 \\
    \hline
    SSIM  &0.8984 & 0.8994 & 0.9005 & 0.9007  & 0.9004 \\
    \hline
  \end{tabular}
  \label{tab:ab_lf}
\vspace{-0.1cm}
\end{table}}

\begin{figure}
\setlength{\belowcaptionskip}{-0.2cm}
\setlength{\abovecaptionskip}{0.1cm}
  \centering
  \includegraphics[width=1.0\linewidth]{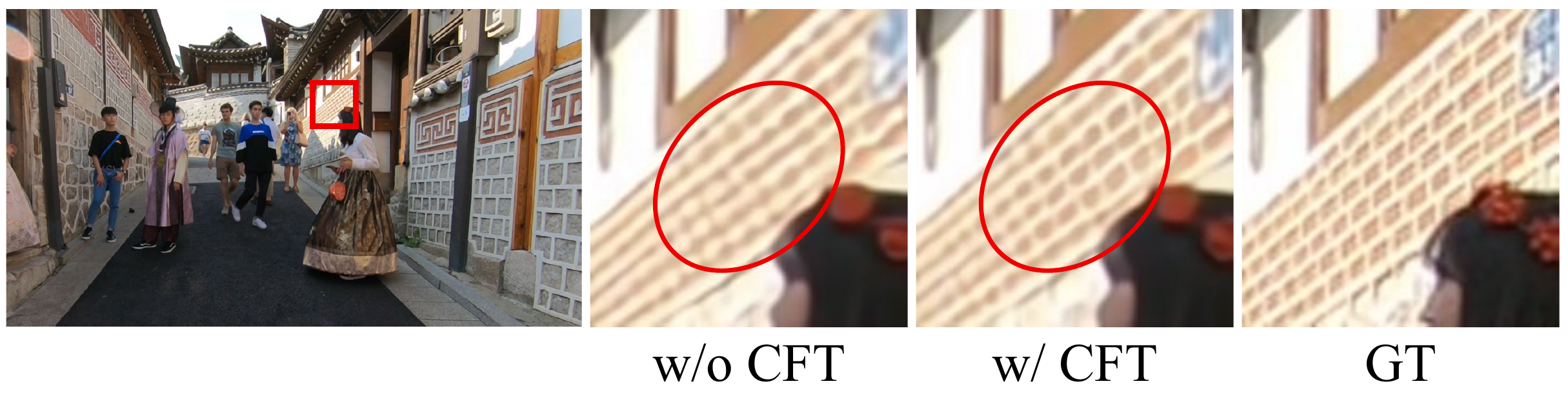}
  \vspace{-0.5cm}
  \caption{Example of without and with the cross-scale feature tokenization (CFT) on the REDS4~\cite{nah2019ntire} dataset. CFT can transfer the clearer textures from larger scales to restore the detailed textures.}
  \label{fig:ab_CEF}
\end{figure}

\noindent\textbf{Cross-scale feature tokenization.} To alleviate the scale-changing problem in sequences, we discuss the impact of token size in the cross-scale feature tokenization (CFT). As shown in Tab.~\ref{tab:CFT}, the first three rows of results show that CFT can extract richer textures as the token scale increases. The performance can improve PSNR from 31.99 to 32.12, indicating that CFT can adapt to scale changes in sequences. In addition, according to the visualizations, as shown in Fig.~\ref{fig:ab_CEF}, cross-scale feature tokenization can introduce finer textures from a larger scale, avoiding the loss of textures caused by scale-changing in long-range sequences. It is also observed that using the larger scale (e.g., $12$) leads to undesirable results. This is because oversized tokens are not conducive to textures learning. In our model, we choose $4$, $6$, and $8$ scales as the token size in CFT.

\setlength{\tabcolsep}{1.0mm}{
\begin{table}\small
  \caption{Ablation study results of cross-scale feature tokenization (CFT) module on the REDS4~\cite{nah2019ntire} dataset, ``S2" and ``S3" represent extracting features from two and three scales, respectively. TTVSR can be interpreted as ``Base+TG+TA+CFT(S3)".}
  \centering
  \vspace{-0.2cm}
  \begin{tabular}{ l || c || c }
    \hline
     Method   &        Token sizes in CFT      &      PSNR/SSIM    \\
    \hline
    Base+TG+TA         &          4        &    31.99/0.9007   \\
    Base+TG+TA+CFT(S2)    &       4, 6      &     32.08/0.9011   \\
    Base+TG+TA+CFT(S3)    &       4, 6, 8      &     \textbf{32.12}/\textbf{0.9021}   \\
    Base+TG+TA+CFT(S3.1)    &       6, 9, 12      &     31.95/0.9004   \\
    Base+TG+TA+CFT(S3.2)    &       8, 12, 16      &     31.91/0.8991   \\
    \hline    
  \end{tabular}
  \label{tab:CFT}
  \vspace{-0.1cm}
\end{table}
}

\begin{figure}
\setlength{\belowcaptionskip}{-0.2cm}
\setlength{\abovecaptionskip}{0.1cm}
  \centering
  \includegraphics[width=1.0\linewidth]{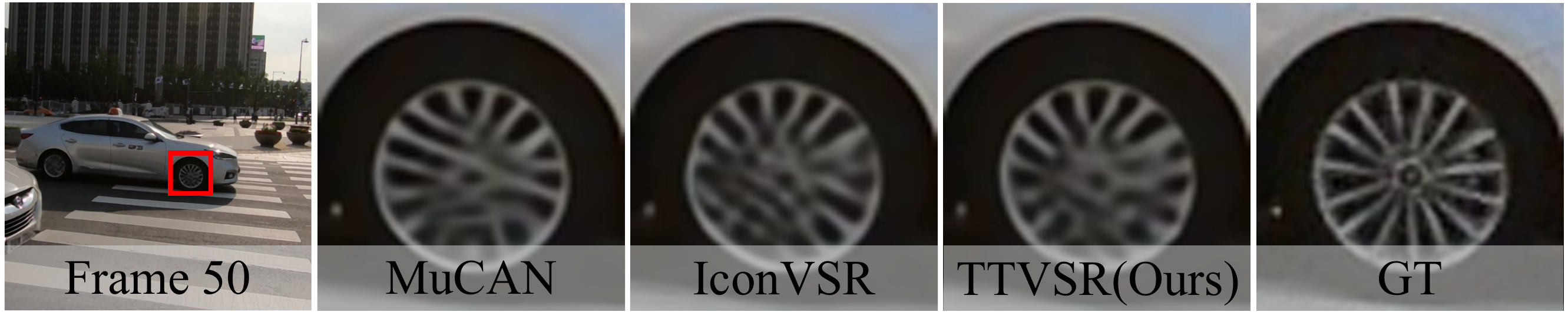}
  \caption{A failure case when rotation occurs.}
  \label{fig:FC}
\vspace{-0.3cm}
\end{figure}

\section{Limitations} 
In this section, we visualize the failure cases of TTVSR in Fig.~\ref{fig:FC}. The motion trajectories are inaccurate when rotation occurs and useful information cannot be transferred through it, thus limiting the performance of our method. However, due to the high difficulty of modeling rotation, other SOTA methods also fail to obtain better performance. It is notable that TTVSR still achieves greater gains than other methods through its powerful long-range modeling ability. More analyses can be found in the supplementary.


\section{Conclusion}
\label{sec:conclusion}
In this paper, we study video super-resolution by leveraging long-range frame dependencies. In particular, we propose a novel trajectory-aware Transformer (TTVSR), which is one of the first works to introduce Transformer architectures in video super-resolution tasks. Specifically, we formulate video frames into pre-aligned trajectories of visual tokens, and calculate attention along trajectories. To implement such formulations, we propose a novel location map to record trajectories, and the location map can online update efficiently by design. TTVSR significantly mitigates computational costs and enables Transformers to model long-range information in videos in an effective way. Experimental results show clear visual margins between the proposed TTVSR and existing SOTA models. In the future, we will focus on 1) evaluating our method in more low-level vision tasks, and 2) extending the trajectory-aware Transformer to high-level vision tasks by more explorations.

\textbf{Acknowledgement.} This work was supported by the NSFC under grants No.61772407. We would like to also thank Tiankai Hang for his help with the paper discussion.

{\small
\bibliographystyle{ieee_fullname}
\bibliography{egbib}
}

\newpage
\begin{appendix}

\section*{Supplementary Material}

In this supplementary material, Sec.~\ref{A} illustrates the details of the algorithm. 
Sec.~\ref{B} and Sec.~\ref{C} describe the detailed architecture and experimental settings, respectively. Sec.~\ref{D} provides the complexity analysis of trajectory-aware attention in TTVSR. Sec.~\ref{E} analyzes the limitations of TTVSR. Finally, Sec.~\ref{F} shows more comparison results.

\section{Algorithm Details}
\label{A}
\setcounter{table}{0}   
\setcounter{figure}{0}
\setcounter{algorithm}{0}
\renewcommand{\thetable}{A.\arabic{table}}
\renewcommand{\thefigure}{A.\arabic{figure}}
\renewcommand{\thealgorithm}{A.\arabic{algorithm}}

In this section, we illustrate the details of the algorithm. It includes the pseudocode of the entire algorithm, the detailed analyzes of the proposed cross-scale feature tokenization module, and location map updating mechanism in our TTVSR. 

\noindent\textbf{Algorithm pseudocode.} As shown in Alg.~\ref{alg}, we describe our proposed TTVSR in the form of pseudocode. Besides, we follow previous works~\cite{huang2017video,chan2021basicvsr} to adopt a bidirectional propagation scheme, where the features in different frames can be propagated backward and forward. For clarity, in this algorithm, we only show the process of forward propagation, and the process of backward propagation is similar to forward propagation.

\begin{figure}[!t]
  \centering
  \includegraphics[width=1.0\linewidth]{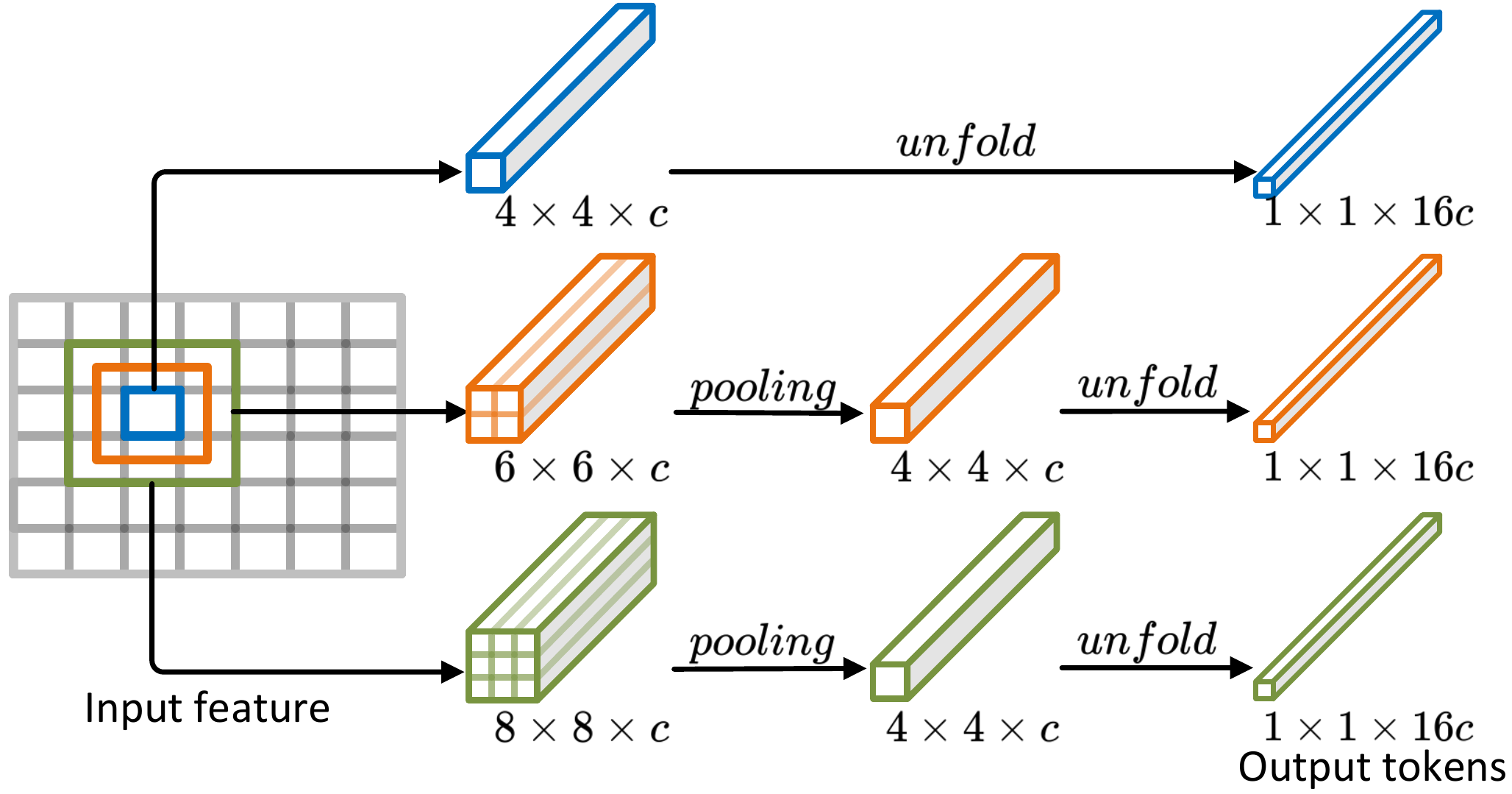}
  \vspace{-0.7cm}
  \caption{An illustration of the cross-scale feature tokenization.}
  \label{fig:cft}
  \vspace{-0.5cm}
\end{figure}

\noindent\textbf{Cross-scale feature tokenization.} As discussed in the main paper, in addition to the more complex motion in the long-range sequence, at the same time, the contents in sequences have the distinct scale changing as it moves. The use of texture from a larger scale often helps to recover more detailed texture on a smaller scale. Therefore, to adapt to the changes of different scales, we propose the cross-scale feature tokenization before trajectory-aware attention to obtain the unified features from the multi-scale. As shown in Fig.~\ref{fig:cft}, we first use the successive unfold and fold operations to expand the receptive field of features, which capture features with more textures on a larger scale. Then, features of different scales are sampled to the same one by a pooling operation, which is used to unify features from different scales. Finally, the features are split by unfolding operation to obtain the output tokens. It is noteworthy that this process can extract features of the whole temporal sequence in parallel, and obtain features of larger scales without adding additional parameters unlike other tasks~\cite{lin2017feature,zhao2017pyramid}.

\noindent\textbf{Location map updating.} As discussed in the main paper, the location maps will change over time. We denotes the updated location maps as ${}^*\!\mathcal{L}^{t}$ and provides a more detailed and formulated description in this section.

When changing from time $T$ to time $T+1$, first, based on Equ. 7 in the main paper, ${}^*\!\mathcal{L}_{m,n}^{T+1}$ represents the coordinate at time $T+1$ of a trajectory which is ended at $(m, n)$ at time $T+1$, which can be expressed as:
\begin{equation}
{}^*\!\mathcal{L}_{m,n}^{T+1}=(m,n).
\end{equation}

\begin{algorithm}[!t]\small
  \caption{The detailed algorithm describe of TTVSR.} 
  \label{alg}
  \begin{algorithmic}[1]
    \Require 
      $\mathbf{I}_{LR}$: $\{I_{LR}^{t}, t \in [1,T]\}$;
      $T$: the length of sequence;
      $\mathcal{L}^{init}$: initialization by uniform discretization;
      $H$ and $W$: the height and width of the feature maps;
      $\phi(\cdot)$ and $\varphi(\cdot)$: embedding networks;
      $\text{S}(\cdot)$: spatial sampling operation;
      $\text{R}(\cdot)$: reconstruction network.
      $\text{P}(\cdot)$: pixel-shuffle.
      $\text{U}(\cdot)$: upsampling operation.
      $\text{H}(\cdot)$: motion estimation network with parameter $\theta$ and an average pooling operation.
      $\text{A}_{traj}(\cdot)$: trajectory-aware attention.
    \Ensure
      $\mathbf{I}_{SR}$: $\{I_{SR}^{t}, t \in [1,T]\}$;\\
      $\mathcal{V}=\{\}$;\\
      $\mathcal{L}=\{\}$;
      \For{$t = 1$; $t<=T$; $t++$}
          \State $O^{t} = \text{H}(I_{LR}^{t},I_{LR}^{t-1};\theta)$;
          \State $\mathcal{L}=\text{S}(\mathcal{L}, O^{t})$;
          \State $\mathcal{L}$ add $\mathcal{L}^{init}$;
          \State $\mathcal{Q}=\phi(I_{LR}^t)=\{q_{\mathcal{L}_{m,n}^{t}}, m\in [1,H], n\in [1,W]\}$;
          \State $\mathcal{K}=\phi(\mathbf{I}_{LR})=\{k_{\mathcal{L}_{m,n}^{t'}}, m\in [1,H], n\in [1,W], t‘\in [1, t-1]\}$;
          \State $\mathcal{V}=\varphi(\mathbf{I}_{LR})=\{v_{\mathcal{L}_{m,n}^{t'}}, m\in [1,H], n\in [1,W], t’\in [1, t-1]\}$;
          \State $F_{atten} = \text{R}(\mathop{\text{A}_{traj}}_{t,m,n}(q_{\mathcal{L}_{m,n}^{t}},k_{\mathcal{L}_{m,n}^{t'}},v_{\mathcal{L}_{m,n}^{t'}}))$ 
          \State $\mathcal{V}$ add $F_{atten}$ (the process of obtaining $F_{atten}$ can be represented as $\varphi$);
          \State $I_{SR}^t = \text{P}(F_{atten})+\text{U}(I_{LR}^t)$
      \EndFor
  \end{algorithmic}
\end{algorithm}

Then we update the existing location maps $\{\mathcal{L}_{m,n}^{1},\cdots,\mathcal{L}_{m,n}^{T}\}$. To build the connection of trajectories between time $T$ and time $T+1$, we introduce a lightweight motion estimation network which computes a backward flow $O^{T+1}$ from $I_{LR}^{T+1}$ to $I_{LR}^{T}$. This process can be formualted as:
\begin{equation}
O^{T+1} = \text{H}(I_{LR}^{T+1},I_{LR}^{T};\theta), 
\end{equation}
where $\text{H}(\cdot)$ is composed of the motion estimation network with parameter $\theta$ and an average pooling operation. The average pooling is used to ensure that the output of the motion estimation is the same size as $\mathcal{L}^{t}$.

Finally, we get the updated coordinates in location map $\mathcal{L}^{t}$ by interpolating between its adjacent coordinates:
\begin{equation}
{}^*\!\mathcal{L}^{t} = \text{S}(\mathcal{L}^{t}, O^{T+1}), 
\label{motion}
\end{equation}
where $\text{S}(\cdot)$ represents the spatial sampling operation on matrix $\mathcal{L}^{t}$ by spatial correlation $O^{T+1}$ (i.e., $grid\_sample$ in PyTorch). Thus far, we have all the location maps for time $T+1$.

\section{Details of Architecture and Runtimes}
\setcounter{table}{0}   
\setcounter{figure}{0}
\setcounter{algorithm}{0}
\renewcommand{\thetable}{B.\arabic{table}}
\renewcommand{\thefigure}{B.\arabic{figure}}
\renewcommand{\thealgorithm}{B.\arabic{algorithm}}

In this section, we will illustrate the detailed architecture and runtimes of TTVSR.
\label{B}

\noindent\textbf{Architecture.} We follow IconVSR~\cite{chan2021basicvsr}, the feature extraction network uses five residual blocks, which are part of embedding operation $\phi(\cdot)$. The feature reconstruction network $\text{R}(\cdot)$ uses 60 residual blocks, which are part of embedding operation $\varphi(\cdot)$. The channel number of feature is set to 64. 


\begin{table}[!t]\small
  \caption{Model Params and Runtime analysis.}
  \centering
  \begin{tabular}{ l || c || c }
    \hline
     Method   &  \#Params  &  Runtime    \\
     \hline
     \quad Flow Estimator  &  1.4M   &  11ms   \\
     \quad Feature Extraction     &  0.4M   &  3ms   \\
     \quad Cross-scale Feature Tokenization     &  0.0M   &  8ms   \\
     \quad Trajectory-aware Attention     &  0.1M   &  114ms   \\
     \quad Reconstruction Network  &  4.8M   &  72ms   \\
    TTVSR Total  &      6.7M    &  203ms   \\
    \hline  
    MuCAN\cite{li2020mucan}  &      13.6M    &  1,202ms   \\
    \hline 
    BasicVSR\cite{chan2021basicvsr}  &      6.3M    &  63ms   \\
    \hline 
    IconVSR\cite{chan2021basicvsr}  &      8.7M    &  70ms   \\
    \hline  
  \end{tabular}
  \label{tab:runtimes}
  \vspace{-0.3cm}
\end{table}

\noindent\textbf{Runtimes.} As shown in Tab.~\ref{tab:runtimes}, we analyze the parameter size and runtime of each component in TTVSR. The runtime is computed on one LR frame with the size of $180 \times320$  and $\times 4$ upsampling, and all models are conducted on an NVIDIA Tesla V100 GPU. 

As shown in Tab.~\ref{tab:runtimes}, our method is much lighter and faster than other attention-based methods (e.g., MuCAN~\cite{li2020mucan}, which is the SOTA attention-based method), this is thanks to our efficient design of trajectory-aware attention. 
Compare with other methods without attention mechanisms, TTVSR achieves higher results with the smaller parameters size and comparable FLOPs (as shown in main paper Table 3).
While it's worth noting that TTVSR is slower than IconVSR~\cite{chan2021basicvsr}. This is because the runtime highly depends on the code implementation and hardware platforms. Specific to our case, the attention mechanism contains a lot of small matrix multiplications which limits our method to run at the high efficiency (as shown in the fourth row of Tab.~\ref{tab:runtimes}, the time used to calculate attention is more than half of the total time). However, we believe that with the optimization of code and hardware, TTVSR can achieve the comparable runtime to IconVSR ~\cite{chan2021basicvsr} as demonstrated by FLOPs which is independent of hardware.

\section{More Experimental Settings}

\setcounter{table}{0}   
\setcounter{figure}{0}
\setcounter{algorithm}{0}
\renewcommand{\thetable}{C.\arabic{table}}
\renewcommand{\thefigure}{C.\arabic{figure}}
\renewcommand{\thealgorithm}{C.\arabic{algorithm}}

\label{C}
\noindent\textbf{Datasets.} We use the \textbf{REDS}\cite{nah2019ntire} and \textbf{Vimeo-90K} \cite{xue2019video} datasets to construct our training data. 
For \textbf{REDS}~\cite{nah2019ntire}, we apply the MATLAB bicubic downsample (BI) degradation on \textbf{REDS4}~\cite{nah2019ntire} to evaluate TTVSR.
For \textbf{Vimeo-90K}~\cite{xue2019video}, we use the Gaussian filter with a standard deviation of $\sigma=1.6$ and downsampling (BD) degradation, and use \textbf{Vid4} \cite{liu2013bayesian}, \textbf{UDM10}~\cite{yi2019progressive} and \textbf{Vimeo-90K-T} \cite{xue2019video} as test sets along with it.

\setlength{\tabcolsep}{1.0mm}{
\begin{table*}[!t]\small
  \caption{Quantitative comparison (PSNR$\uparrow$, SSIM$\uparrow$, and LPIPS$\downarrow$) on the REDS4~\cite{nah2019ntire} dataset for $4\times$ video super-resolution. The results are tested on RGB channels. \textcolor{red}{Red} indicates the best and \textcolor{blue}{blue} indicates the second best performance (best view in color).}
  \centering
  \begin{tabular}{ l  || c  c  c  c  c  c  c  c  c  c}
    \hline
    Method    &  Bicubic & RCAN~\cite{zhang2018image} & CSNLN~\cite{mei2020image} & TOFlow~\cite{xue2019video} & DUF~\cite{jo2018deep} &  EDVR~\cite{wang2019edvr}  &  MuCAN~\cite{li2020mucan}  & BasicVSR~\cite{chan2021basicvsr} & IconVSR~\cite{chan2021basicvsr} & \textbf{TTVSR} \\
    \hline
    \hline
    PSNR   &  26.14 & 28.78 & 28.83 & 27.98 & 28.63 &   31.09  &  30.88  &   31.42 &     \textcolor{blue}{31.67}  & \textcolor{red}{32.12}\\
    \hline
    SSIM & 0.7292  &  0.8200  &  0.8196  &  0.7990  &  0.8251   &  0.8800  &  0.8750  &   0.8909 &    \textcolor{blue}{0.8948}  & \textcolor{red}{0.9021}\\
    \hline
    LPIPS  &  0.3395  &  0.2716  &  0.2668  &  0.2969  &  0.2911    &  0.2225  &  0.2112  &   0.1979 &     \textcolor{blue}{0.1890}  & \textcolor{red}{0.1786}\\
    \hline    
    
  \end{tabular}
  \label{tab:LPIPS}
\end{table*}
}

\noindent\textbf{Settings.}
To leverage the information of the whole sequence, we follow previous works~\cite{huang2017video,chan2021basicvsr} to adopt a bidirectional propagation scheme. In the process of bidirectional propagation, the same parameters are shared, and the final feature at each frame is obtained by cascading bidirectional output features. 

For \textbf{REDS}~\cite{nah2019ntire}, we use sequences with a length of 50 as inputs, and loss is computed for the 50 output frames. 
For \textbf{Vimeo-90K}~\cite{xue2019video}, we augment the sequence by flipping twice to extend the length of the sequence to 28 as input, and the inference results on \textbf{Vimeo-90K-T}~\cite{xue2019video} are the average of output frames derived from the same frame extension.

The Charbonnier penalty loss~\cite{lai2017deep} is applied on whole sequence between the ground-truth and restored SR frame. It can be formualted as:
\begin{equation}
\ell=\frac{1}{T}\sum_{t=1}^{T}\sqrt{\|I_{HR}^t-I_{SR}^t\|^{2}+\varepsilon^2}, 
\end{equation}
where $\varepsilon=1 \times 10^{-8}$. $T$ denotes the sequence length.
All experiments are conducted on a server with Python 3.9, PyTorch 1.9, and $8\times$NVIDIA Tesla V100 GPUs.

\section{Complexity Analysis}
\label{D}
\setcounter{table}{0}   
\setcounter{figure}{0}
\setcounter{algorithm}{0}
\renewcommand{\thetable}{D.\arabic{table}}
\renewcommand{\thefigure}{D.\arabic{figure}}
\renewcommand{\thealgorithm}{D.\arabic{algorithm}}

By introducing trajectories into Transformer in our proposed TTVSR, the attention calculation on $\mathcal{K}$ and $\mathcal{V}$ can be significantly reduced because it can avoid the computation on spatial dimension compared with vanilla vision Transformers. In this section, we analyze the complexity of trajectory-aware attention calculation in detail. 

Take the attention process of one token in $\mathcal{Q}$ as an example, when a sequence of length $T$ and size $C\cdot H\cdot W$ is input as $\mathcal{K}$, the size of tokens and the number of tokens can be expressed as $C\cdot D_{h}\cdot D_{w}$ and $T\cdot \frac{H}{D_{h}}\cdot \frac{W}{D_{w}}$, respectively. The similarity of attention mechanisms in vanilla vision Transformer has a computational cost of:
\begin{equation}
(T\cdot \frac{H}{D_{h}}\cdot \frac{W}{D_{w}}) \cdot (C\cdot D_{h}\cdot D_{w}).
\label{equ:1}
\end{equation}

The similarity of trajectory-aware attention mechanisms in TTVSR has a computational cost of:
\begin{equation}
(T\cdot 1\cdot 1) \cdot (C\cdot D_{h}\cdot D_{w}),
\label{equ:2}
\end{equation}
here, the attention calculation on spatial dimension is avoided, so the number of tokens in the attention calculation is reduced from $(T\cdot \frac{H}{D_{h}}\cdot \frac{W}{D_{w}})$ to $(T\cdot 1\cdot 1)$.

By expressing attention as the multiplication of tokens, we get a reduction in computation of:
\begin{equation}
\frac{(T\cdot 1\cdot 1) \cdot (C\cdot D_{h}\cdot D_{w})}{(T\cdot \frac{H}{D_{h}}\cdot \frac{W}{D_{w}}) \cdot (C\cdot D_{h}\cdot D_{w})}=\frac{1}{(\frac{H}{D_{h}}\cdot \frac{W}{D_{w}})}.
\label{equ:3}
\end{equation}

In general, by introducing trajectories into Transformer, the computational complexity of attention is reduced by $(\frac{H}{D_{h}}\cdot \frac{W}{D_{w}})$ times, and provides a more efficient way to enable our TTVSR can directly leverage the information from a distant video frame.

\section{Limitation Analysis}

\setcounter{table}{0}   
\setcounter{figure}{0}
\setcounter{algorithm}{0}
\renewcommand{\thetable}{E.\arabic{table}}
\renewcommand{\thefigure}{E.\arabic{figure}}
\renewcommand{\thealgorithm}{E.\arabic{algorithm}}

\label{E}
In this section, we discuss the limitations of TTVSR.

\noindent\textbf{Length of sequence.} The design of TTVSR is to fully utilize the temporal information of more frames in the sequence, so the performance improvement is limited for shorter sequences. (e.g., each sequence contains seven frames on Vimeo-90K~\cite{xue2019video}). Therefore, to enable long-range sequence capability, for \textbf{Vimeo-90K}~\cite{xue2019video}, we augment the sequence by flipping twice to extend the length of the sequence to 28 as input. However, due to the limitation of original sequences, the improvement is still limited. We will add relevant features on spatial dimensions to further enhance the utilization of spatial information in the model.

\noindent\textbf{Training Time.} To make the model have the ability to model long-range sequences,  we input the sequences as long as possible during training. However, longer sequences often lead to longer training times. Next, we will optimize the training process so that TTVSR can speed up the training process with long sequences input.

\noindent\textbf{GPU memory.} According to the model design, TTVSR needs to store the features of each moment during inferring and training. It inevitably takes up GPU memory. Therefore, the next step is to design an adaptive feature storage mechanism dynamically and selectively retain the useful features for reconstruction.

\section{More Results}

\setcounter{table}{0}   
\setcounter{figure}{0}
\setcounter{algorithm}{0}
\renewcommand{\thetable}{F.\arabic{table}}
\renewcommand{\thefigure}{F.\arabic{figure}}
\renewcommand{\thealgorithm}{F.\arabic{algorithm}}

\label{F}

In this section, to further verify the effectiveness of our method, we compare the perceptual results and show more comparison results among the proposed TTVSR and other advanced methods on four different benchmarks.

\noindent\textbf{Perceptual results.} We use LPIPS~\cite{zhang2018unreasonable} as a widely used metric to evaluate perceptual quality. Results, shown in the following Additional Tab.~\ref{tab:LPIPS}, demonstrate that TTVSR is still highly superior in the perceptual metrics.

\begin{figure*}[!h]
  \centering
  \includegraphics[width=0.9\linewidth,page={1}]{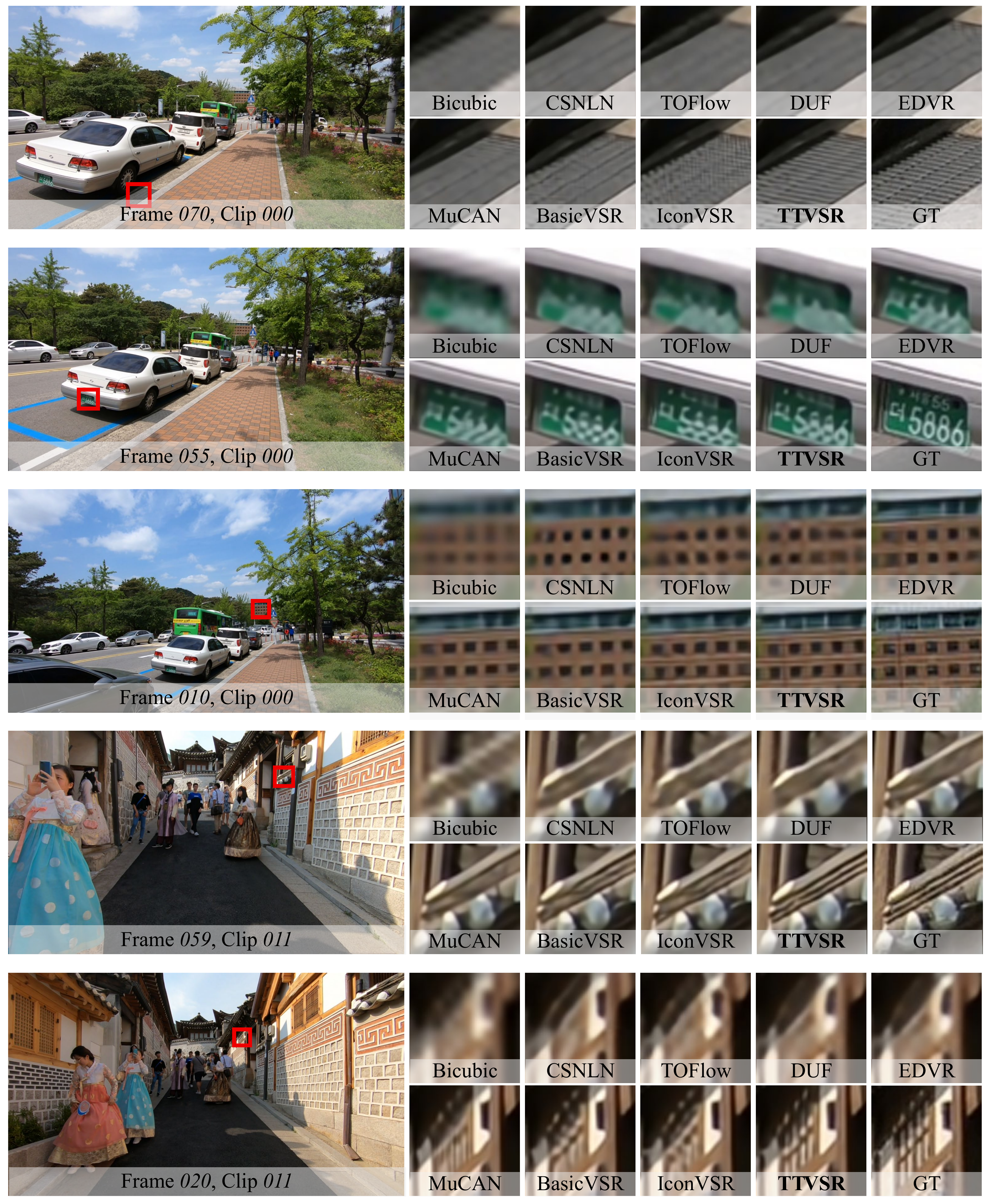}
  \caption{Visual results on REDS4~\cite{nah2019ntire} for $4 \times$ scaling factor. The frame number is shown at the bottom of each case. Zoom in to see better visualization.}
  \label{fig:reds1}
\end{figure*}

\begin{figure*}[!h]
  \centering
  \includegraphics[width=0.9\linewidth,page={2}]{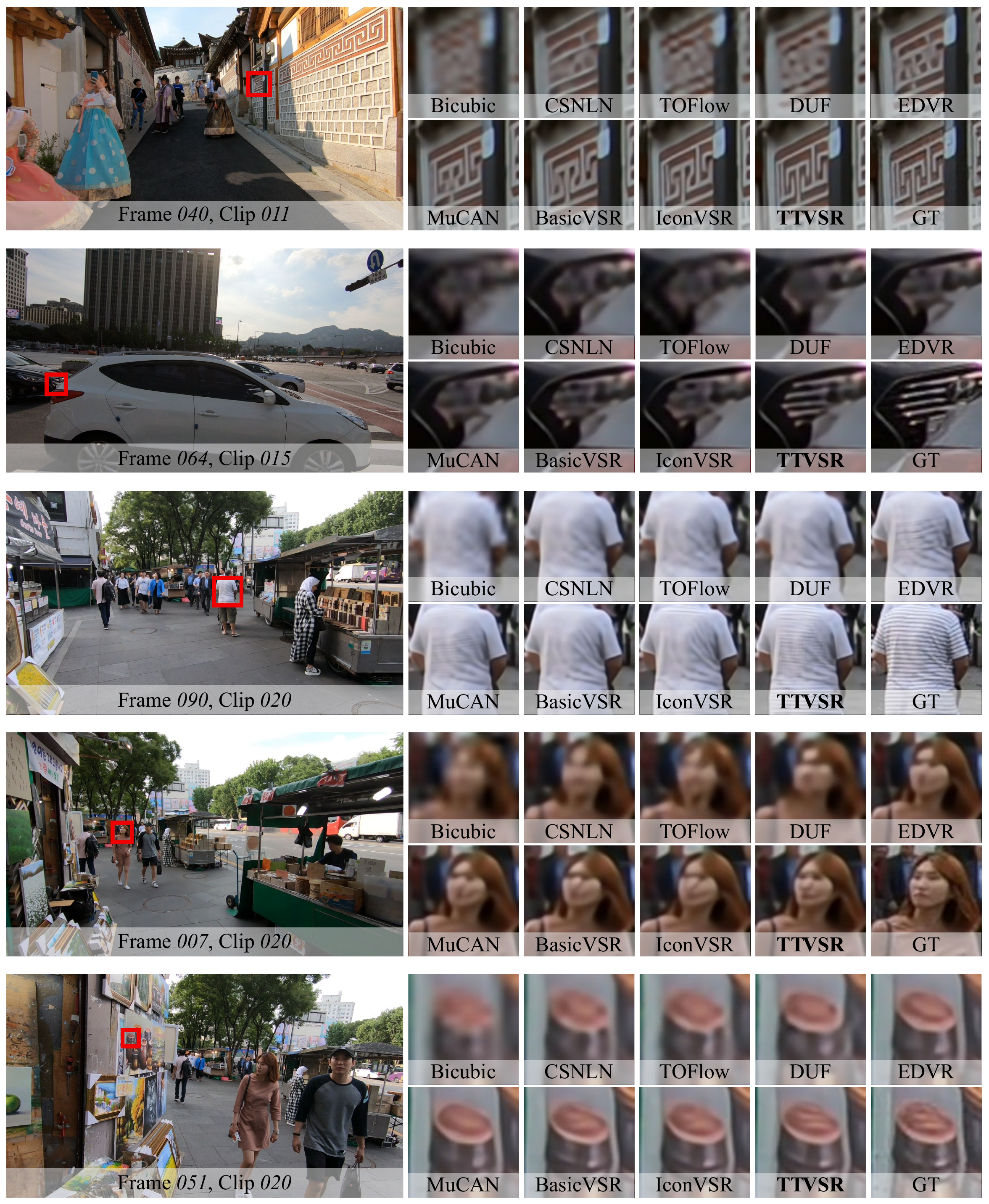}
  \caption{Visual results on REDS4~\cite{nah2019ntire} for $4 \times$ scaling factor. The frame number is shown at the bottom of each case. Zoom in to see better visualization.}
  \label{fig:reds2}
\end{figure*}

\begin{figure*}[!h]
  \centering
  \includegraphics[width=0.8\linewidth,page={3}]{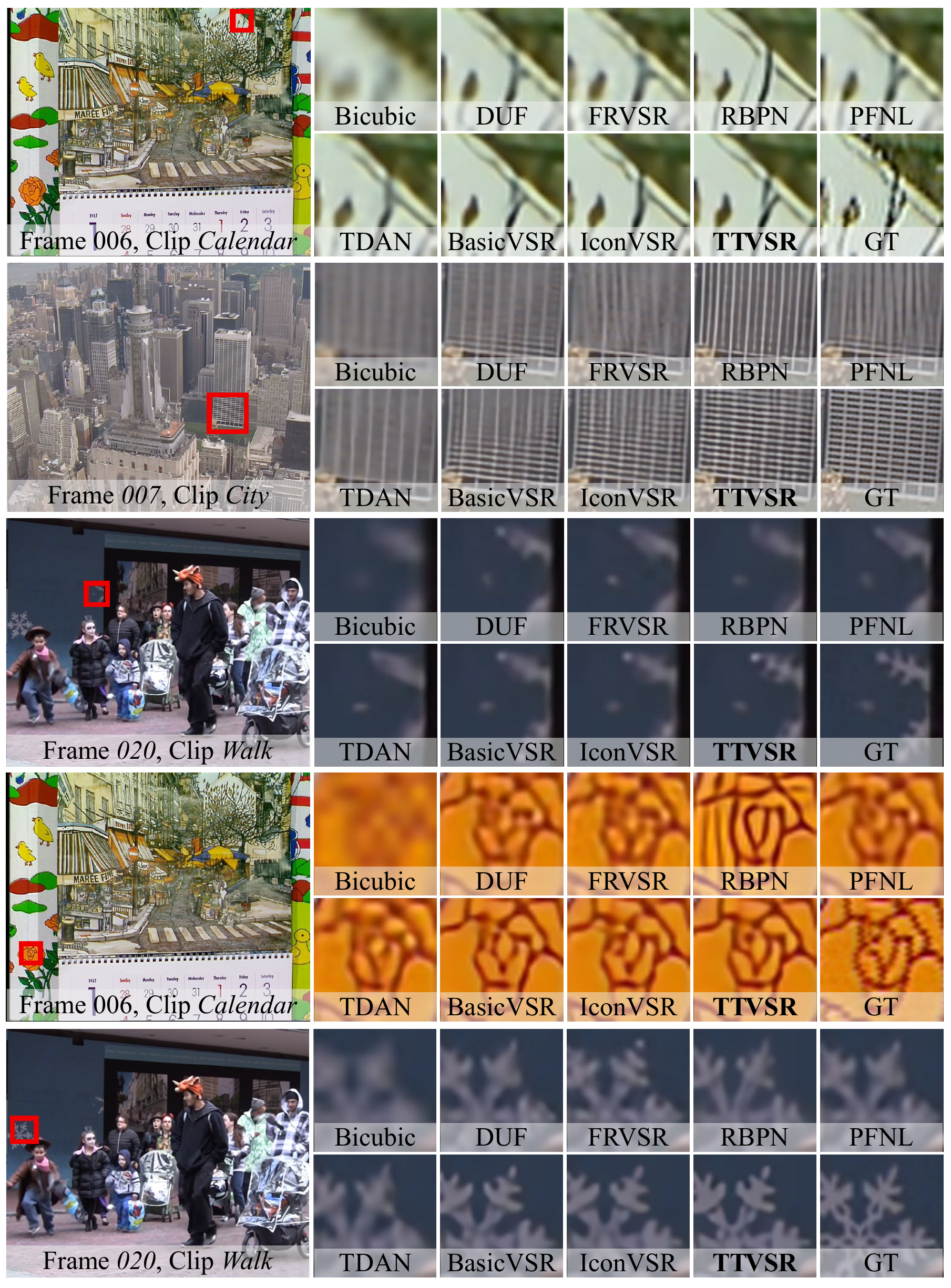}
  \caption{Visual results on Vid4~\cite{liu2013bayesian} for $4 \times$ scaling factor. The frame number is shown at the bottom of each case. Zoom in to see better visualization.}
  \label{fig:vid4}
\end{figure*}

\begin{figure*}[!h]
  \centering
  \includegraphics[width=0.8\linewidth,page={4}]{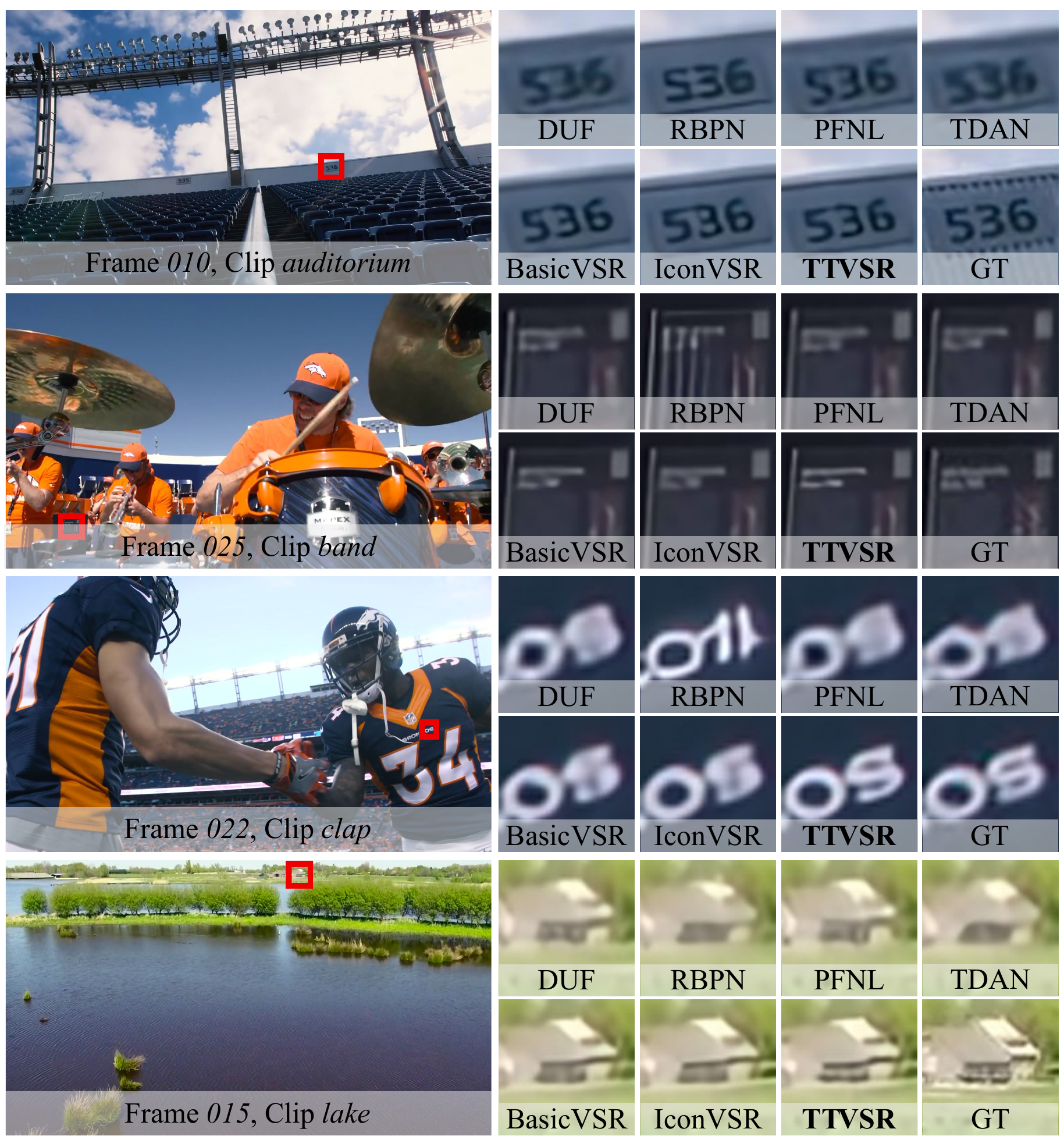}
  \caption{Visual results on UDM10~\cite{yi2019progressive} for $4 \times$ scaling factor. The frame number is shown at the bottom of each case. Zoom in to see better visualization.}
  \label{fig:udm10}
\end{figure*}

\begin{figure*}[!h]
  \centering
  \includegraphics[width=0.9\linewidth,page={5}]{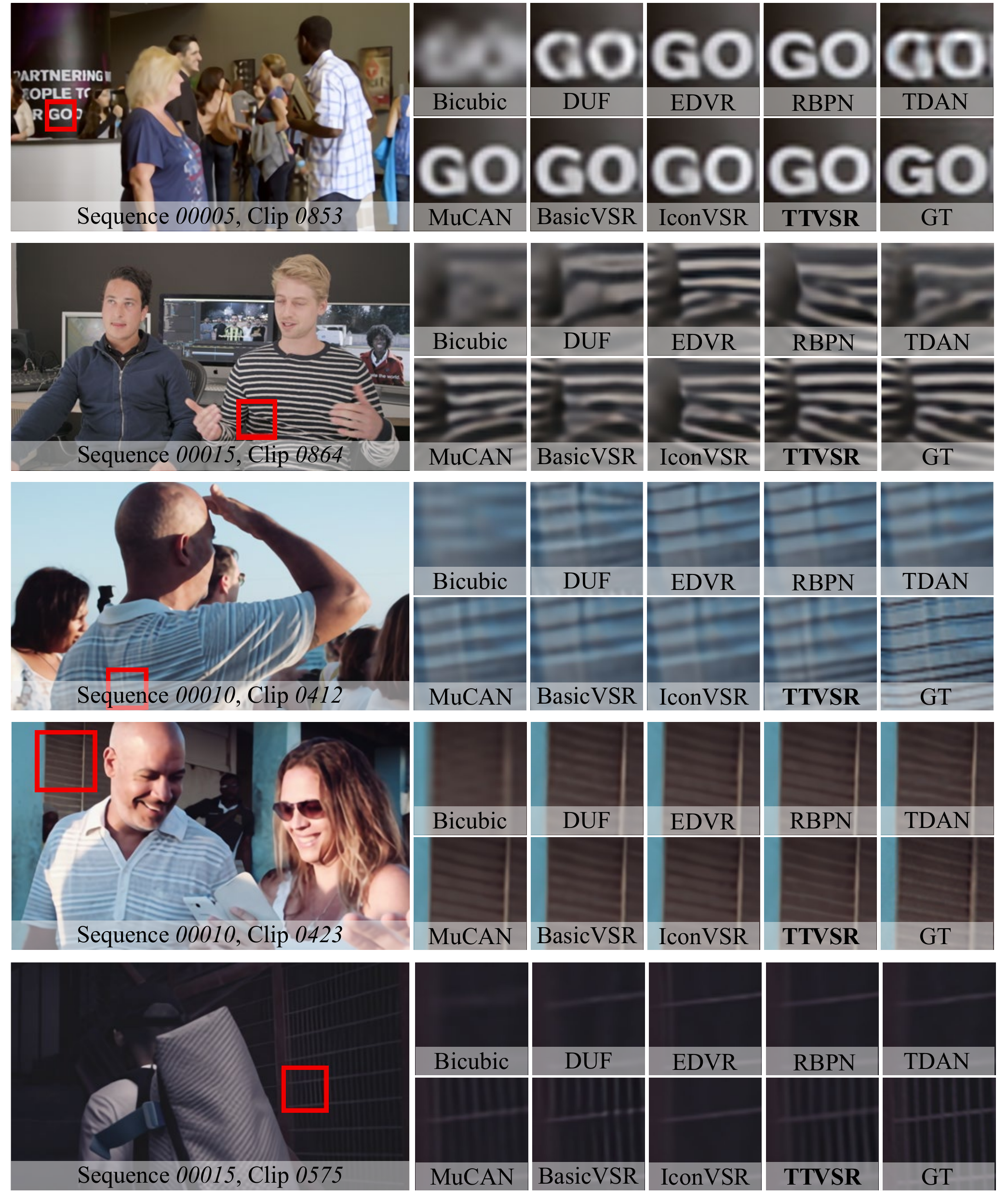}
  \caption{Visual results on Vimeo-90K-T~\cite{xue2019video} for $4 \times$ scaling factor. The frame number is shown at the bottom of each case. Zoom in to see better visualization.}
  \label{fig:vimeo1}
\end{figure*}

\begin{figure*}[!h]
  \centering
  \includegraphics[width=0.9\linewidth,page={6}]{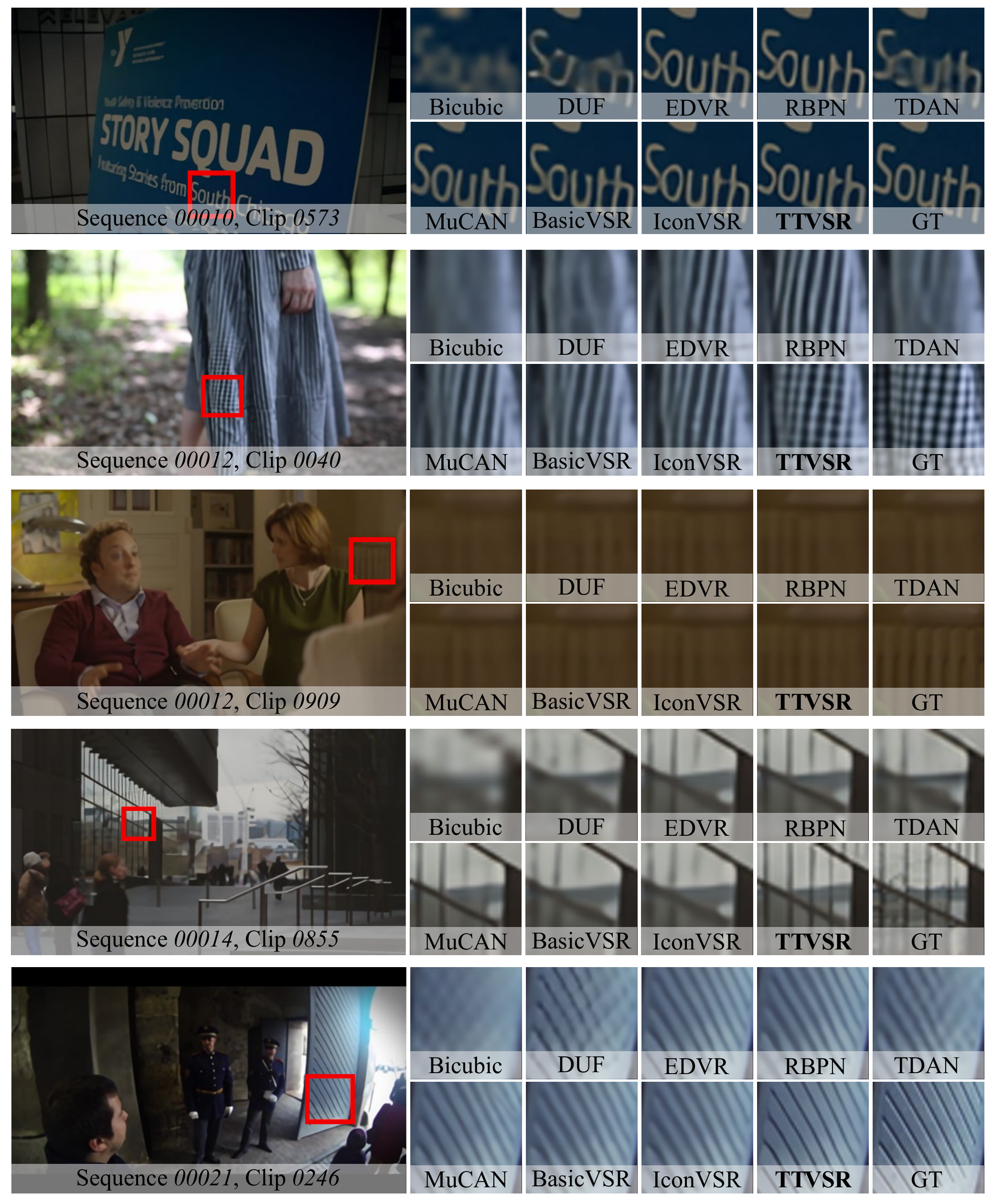}
  \caption{Visual results on Vimeo-90K-T~\cite{xue2019video} for $4 \times$ scaling factor. The frame number is shown at the bottom of each case. Zoom in to see better visualization.}
  \label{fig:vimeo2}
\end{figure*}

\noindent\textbf{Visualization results.} We show more comparison results among the proposed TTVSR and other advanced methods on four different benchmarks. The results on \textbf{REDS4}~\cite{nah2019ntire}, \textbf{Vid4}~\cite{liu2013bayesian}, \textbf{UDM10}~\cite{yi2019progressive} and \textbf{Vimeo-90K-T}~\cite{xue2019video} are shown in Fig.~\ref{fig:reds1}-\ref{fig:reds2}, Fig.~\ref{fig:vid4}, Fig.~\ref{fig:udm10} and Fig.~\ref{fig:vimeo1}-\ref{fig:vimeo2}, respectively.

\clearpage
\clearpage

\end{appendix}
\end{document}